\documentclass[12pt]{article}

\usepackage{amsfonts, amsmath}

\newcommand{\be}{\begin{equation}} \newcommand{\ee}{\end{equation}}

\begin{document}
\title{\bf The Quantum and Statistical Theory of Early
 Universe and Its Possible Applications to Cosmology} \thispagestyle{empty}

\author{Alex.E.Shalyt-Margolin\hspace{1.5mm}\thanks
{Phone (+375) 172 883438; e-mail: alexm@hep.by; a.shalyt@mail.ru}
\thanks{Fax (+375) 172 326075}
}
\date{}
\maketitle
 \vspace{-25pt}
{\footnotesize\noindent Laboratory of the Quantum Field Theory,
National Center of Particles and High Energy Physics, Bogdanovich
Str. 153, Minsk 220040, Belarus\\ {\ttfamily{\footnotesize
 }}

\rm\normalsize \vspace{0.5cm}
\begin{abstract}
In this paper  a new approach to investigation of Quantum and
Statistical Mechanics of the Early Universe (Planck scale) -
density matrix deformation - is proposed. The deformation is
understood as an extension of a particular theory by inclusion of
one or several additional parameters in such a way that the
initial theory appears in the limiting transition.  In the first
part of this chapter Quantum Mechanics of the Early Universe is
treated as a Quantum Mechanics with Fundamental Length considering
the fact that different approaches to quantum gravitation
exhibited in the Early Universe were inevitably leading to the
notion of fundamental length on the order of Planck's. Besides,
this is possible due to the involvement in this theory of the
Generalized Uncertainty Relations. And Quantum Mechanics with
Fundamental Length is obtained as a deformation of Quantum
Mechanics. The distinguishing feature of the proposed approach as
compared to the previous ones is the fact that here the density
matrix is subjected to deformation, rather than commutators or
(that is the same) Heisenberg's Algebra. In this chapter the
density matrix obtained by deformation of the quantum-mechanical
one is referred to as a density pro-matrix. Within our approach
two main features of Quantum Mechanics are conserved: the
probabilistic interpretation of the theory and the well-known
measuring procedure associated with this interpretation. The
proposed approach allows for description of dynamics, in
particular, the explicit form of deformed Liouville equation and
the deformed Shrodinger's picture. Some implications of the
obtained results are discussed including the singularity problem,
hypothesis of cosmic censorship, possible improvement of the
definition for statistical entropy. It is shown that owing to the
obtained results one is enabled to deduce in a simple and natural
way the Bekenstein-Hawking formula for black hole entropy in a
semiclassical approximation. In the second part of the chapter it
is demonstrated that Statistical Mechanics of the Early Universe
is a deformation of the conventional Statistical Mechanics. The
statistical-mechanics deformation is constructed by analogy to the
earlier quantum mechanical results. As previously, the primary
object is a density matrix, but now the statistical one. The
obtained deformed object is referred to as a statistical density
pro-matrix. This object is explicitly described, and it is
demonstrated that there is a complete analogy in the construction
and properties of quantum-mechanics and statistical density
matrices at Plank scale (i.e. density pro-matrices). It is shown
that an ordinary statistical density matrix occurs in the
low-temperature limit at temperatures much lower than the Plank's.
The associated deformation of a canonical Gibbs distribution is
given explicitly. Also consideration is being given to rigorous
substantiation of the Generalized Uncertainty Relations as applied
in thermodynamics. And in the third part of the chapter the
results obtained are applied to solution of the Information
Paradox (Hawking) Problem. It is demonstrated that involvement of
black holes in the suggested approach in the end twice causes
nonunitary transitions resulting in the unitarity. In parallel
this problem is considered in other terms: entropy density,
Heisenberg algebra deformation terms, respective deformations of
Statistical Mechanics, - all showing the identity of the basic
results. From this an explicit solution for Hawking's  informaion
paradox has been derived. Besides, it is shown that owing to the
proposed approach a new small parameter is derived in physics, the
principal features of which are its dimensionless character and
its association with all the fundamental constants. In the last
part of this chapter it is shown that on the basis of the above
parameter the Universe may be considered as nonuniform lattice in
the finite-dimensional hypercube. Besides, possible applications
of the results are proposed.
\end{abstract}

\section{Introduction}
In the last few years the Early Universe has aroused considerable
interest of the researchers. This may be caused by several facts.
First, a Big Bang theory is presently well grounded and has
established experimental status. Second, acknowledged success of
the inflation model and its interface with high-energy physics.
Third, various approaches to topical problems of the fundamental
physics, specifically to the problem of divergence in a quantum
theory or singularity in the General Relativity Theory, in some or
other way lead to the problem of quantum-gravitational effects and
their adequate description. And all the above aspects are related
to the Early Universe. Because of this, investigation of the Early
Universe is of particular importance. The Early Universe is
understood as a Universe at the first Planck's moments following
the Big Bang when energies and scales were on the order of
Planck's.
\\In this chapter a new approach to
investigation of Quantum and Statistical Mechanics of the Early
Universe  - density matrix deformation - is proposed. The
deformation is understood as an extension of a particular theory
by inclusion of one or several additional parameters in such a way
that the initial theory appears in the limiting transition. The
most clear example is QM being a deformation of Classical
Mechanics. The parameter of deformation in this case is the
Planck's constant $\hbar$. When $\hbar\rightarrow 0$ QM goes to
Classical Mechanics.
\\In the first part of this chapter Quantum Mechanics of the
Early Universe is treated as a Quantum Mechanics with Fundamental
Length. This becomes possible since different approaches to
quantum gravitation exhibited in the Early Universe unavoidably
involve the notion of fundamental length on the order of Planck's
(see \cite{r1} and the references). Also this is possible due to
the involvement in this theory of the Generalized Uncertainty
Relations. And Quantum Mechanics with Fundamental Length is
obtained as a deformation of Quantum Mechanics. The distinguishing
feature of the proposed approach as compared to the previous ones
is the fact that here the density matrix is subjected to
deformation, rather than commutators or (that is the same)
Heisenberg's Algebra. In this chapter the density matrix obtained
by deformation of the quantum-mechanical one is referred to as a
density pro-matrix. Within our approach two main features of
Quantum Mechanics are conserved: the probabilistic interpretation
of the theory and the well-known measuring procedure associated
with this interpretation. The proposed approach allows for
description of dynamics, in particular, the explicit form of
deformed Liouville equation and the deformed Shrodinger's picture.
Some implications of the obtained results are discussed including
the singularity problem, hypothesis of cosmic censorship, possible
improvement of the definition for statistical entropy. It is shown
that owing to the obtained results one is enabled to deduce in a
simple and natural way the Bekenstein-Hawking formula for black
hole entropy in a semiclassical approximation. In the second part
of the chapter it is demonstrated that Statistical Mechanics of
the Early Universe is a deformation of the conventional
Statistical Mechanics. The statistical-mechanics deformation is
constructed by analogy to the earlier quantum mechanical results.
As previously, the primary object is a density matrix, but now the
statistical one. The obtained deformed object is referred to as a
statistical density pro-matrix. This object is explicitly
described, and it is demonstrated that there is a complete analogy
in the construction and properties of quantum-mechanics and
statistical density matrices at Plank scale (i.e. density
pro-matrices). It is shown that an ordinary statistical density
matrix occurs in the low-temperature limit at temperatures much
lower than the Plank's. The associated deformation of a canonical
Gibbs distribution is given explicitly. Also consideration is
being given to rigorous substantiation of the Generalized
Uncertainty Relations as applied in thermodynamics. And in the
third part of the chapter the results obtained are applied to
solution of the Information Paradox (Hawking) Problem. It is
demonstrated that involvement of black holes in the suggested
approach in the end twice causes nonunitary transitions resulting
in the unitarity. In parallel this problem is considered in other
terms: entropy density, Heisenberg algebra deformation terms,
respective deformations of Statistical Mechanics, - all showing
the identity of the basic results. From this an explicit solution
for Hawking's  Informaion paradox has been derived. Besides, it is
shown that owing to the proposed approach a new small parameter is
derived in physics, the principal features of which are its
dimensionless character and its association with all the
fundamental constants. In the last part of the chapter it is shown
that on the basis of the above parameter the Universe may be
considered as nonuniform lattice in the finite-dimensional
hypercube. Besides, possible applications of the results are
proposed.
\\ {\bf This chapter is devoted by the author to two anniversaries, namely:
the 70-th anniversary of Academician Ludvig Dmitrievich Faddeev,
Russian Academy Sciences, whose work \cite{Fadd} and report at the
round-table Conference of the XI International Congress on
Mathematical Physics in Paris, July 1994, became a "guiding star"
for the author in his research; and the 60-th anniversary of my
first science manager Professor Vassilii Ivanovich Strazhev,
presently Rector of the Belarusian State University.}

\section{Fundamental Length and Density Matrix}

Using different approaches (String Theory \cite{r2}, Gravitation
\cite{r3}, etc.), the authors of numerous papers issued over the
last 14-15 years have pointed out that Heisenberg's Uncertainty
Relations should be modified. Specifically, a high energy
correction has to appear
\begin{equation}\label{U2}
\triangle x\geq\frac{\hbar}{\triangle p}+\alpha^{\prime}
L_{p}^2\frac{\triangle p}{\hbar}.
\end{equation}
\noindent Here $L_{p}$ is the Planck's length:
$L_{p}=\sqrt\frac{G\hbar}{c^3}\simeq1,6\;10^{-35}\;m$ and
 $\alpha^{\prime} > 0$ is a constant. In \cite{r3} it was shown
that this constant may be chosen equal to 1. However, here we will
use $\alpha^{\prime}$ as an arbitrary constant without giving it
any definite value. Equation (\ref{U2})  is identified as the
Generalized Uncertainty Relations in Quantum Mechanics.
\\The inequality (\ref{U2}) is quadratic in $\triangle p$:
\begin{equation}\label{U3}
\alpha^{\prime} L_{p}^2({\triangle p})^2-\hbar \triangle x
\triangle p+ \hbar^2 \leq0,
\end{equation}
from whence the fundamental length is
\begin{equation}\label{U4}
\triangle x_{min}=2\sqrt\alpha^{\prime} L_{p}.
\end{equation}
Since in what follows we proceed only from the existence of
fundamental length, it should be noted that this fact was
established apart from GUR as well. For instance, from an ideal
experiment associated with Gravitational Field and Quantum
Mechanics a lower bound on minimal length was obtained in
\cite{r7}, \cite{r8} and  improved in \cite{r9} without using GUR
to an estimate of the form $\sim L_{p}$. As reviewed previously in
\cite{r1}, the fundamental length appears quite naturally at
Planck scale, being related to the quantum-gravitational
effects.\noindent Let us  consider equation (\ref{U4}) in some
detail.  Squaring both its sides, we obtain
\begin{equation}\label{U5}
(\overline{\Delta\widehat{X}^{2}})\geq 4\alpha^{\prime} L_{p}^{2},
\end{equation}
Or in terms of density matrix
\begin{equation}\label{U6}
Sp[(\rho \widehat{X}^2)-Sp^2(\rho \widehat{X}) ]\geq
4\alpha^{\prime} L_{p}^{2 }=l^{2}_{min}>0,
\end{equation}
where $\widehat{X}$ is the coordinate operator. Expression
(\ref{U6}) gives the measuring rule used in QM.As distinct from
QM,however, in the are considered here the right-hand side of
(\ref{U6}) can not be brought arbitrary close to zero as it is
limited by $l^{2}_{min}>0$, where because of GUR $l_{min} \sim
L_{p}$.
\\Apparently, this may be due to the fact that QMFL  is
unitary non-equivalent to  QM. Actually, in QM the left-hand side
of (\ref{U6}) can be chosen arbitrary close to zero, whereas in
QMFL this is impossible. But if two theories are unitary
equivalent then the form of their spurs should be retained.
Besides, a more important aspect is contributing to unitary
non-equivalence of these two theories: QMFL contains three
fundamental constants (independent parameters) $G$, $c$ and
$\hbar$, whereas QM contains only one: $\hbar$. Within an
inflationary model (see \cite{r10}), QM is the low-energy limit of
QMFL (QMFL turns to QM) for the expansion of the Universe.This is
identical for all cases of transition from Planck's energies to
the normal ones \cite{r1}. In special case of using GUR, the
second term in the right-hand side of (\ref{U2}) vanishes and GUR
turn to UR \cite{r6}. A natural way for studying QMFL is to
consider this theory as a deformation of QM, turning to QM at the
low energy limit (during expansion of the Universe after the Big
Bang). We will consider precisely this option.In this paper,
unlike the works of other authors (e.g. see \cite{r5}) the density
matrix is deformed rather than commutators,whereas the fundamental
fundamental quantum-mechanical measuring rule (\ref{U6}) is left
without changes. Here the following question may be formulated:
how should be deformed density matrix conserving
quantum-mechanical measuring rules in order to obtain
self-consistent measuring procedure in QMFL? To answer this
question, we use the R-procedure. First consider R-procedure both
at the Planck's and  low-energy scales. At the Planck's scale $a
\approx il_{min}$ or $a \sim iL_{p}$, where $i$ is a small
quantity. Further $a$ tends to infinity and we obtain for density
matrix \cite{shalyt1}-\cite{shalyt5}:
 $$Sp[\rho a^{2}]-Sp[\rho
a]Sp[\rho a] \simeq l^{2}_{min}\;\; or\;\; Sp[\rho]-Sp^{2}[\rho]
\simeq l^{2}_{min}/a^{2}.$$

 Therefore:

 \begin{enumerate}
 \item When $a < \infty$, $Sp[\rho] =
Sp[\rho(a)]$ and
 $Sp[\rho]-Sp^{2}[\rho]>0$. Then  \newline $Sp[\rho]<1$
 that corresponds to the QMFL case.
\item When $a = \infty$, $Sp[\rho]$ does not depend on $a$ and
$Sp[\rho]-Sp^{2}[\rho]\rightarrow 0$. Then  $Sp[\rho]=1$ that
corresponds to the QM case.
\end{enumerate}
Interesting,how should be  interpreted 1 and 2 ? Does  the above
analysis agree with the main result from \cite{r30} \footnote
{"... there cannot be any physical state which is a position
eigenstate since an eigenstate would of course have zero
uncertainty in position"}? Note the agreement is well. Indeed, any
time when the state vector reduction (R-procedure) place in QM
always an eigenstate (value) is chosen exactly. In other words,
the probability is equal to 1. As it was pointed out in statement
1, the situation changes when we consider QMFL: it is impossible
to measure coordinates exactly,they never will be absolutely
reliable. In all cases  we obtain a probability less than 1
($Sp[\rho]=p<1$). In other words, any R-procedure in QMFL leads to
an eigenvalue, but only with a probability less than 1. This
probability is as near to 1 as far the difference between
measuring scale $a$ and $l_{min}$ is growing, or in other words,
the second term in (\ref{U2}) becomes insignificant and we turn to
QM. Here there is no  contradiction with \cite{r30}. In QMFL there
are no exact coordinate eigenstates (values) as well as there are
no pure states. In this paper we consider not the operator
properties in QMFL as it was done in \cite{r30} but density-matrix
properties.

 The  properties of density matrix in
QMFL and QM have to be different. The only reasoning in this case
may be as follows: QMFL must differ from QM, but in such a way
that in the low-energy limit a density matrix in QMFL be
coincident  with the density matrix in QM. That is to say, QMFL is
a deformation of QM and the parameter of deformation depends on
the measuring scale. This means that in QMFL $\rho=\rho(x)$, where
$x$ is the scale, and for $x\rightarrow\infty$  $\rho(x)
\rightarrow \widehat{\rho}$, where $\widehat{\rho}$ is the density
matrix in QM.

Since on the Planck's scale $Sp[\rho]<1$, then for such scales
$\rho=\rho(x)$, where $x$ is the scale, is not a density matrix as
it is generally defined in QM. On Planck's scale $\rho(x)$ is
referred to as  "density pro-matrix". As follows from the above,
the density matrix $\widehat{\rho}$ appears in the limit
\cite{shalyt1}-\cite{shalyt5}:
\begin{equation}\label{U12}
\lim\limits_{x\rightarrow\infty}\rho(x)\rightarrow\widehat{\rho},
\end{equation}
when  QMFL turns to QM.

Thus, on Planck's scale the density matrix is inadequate to obtain
all information about the mean values of operators. A "deformed"
density matrix (or pro-matrix) $\rho(x)$ with $Sp[\rho]<1$ has to
be introduced because a missing part of information $1-Sp[\rho]$
is encoded in the quantity $l^{2}_{min}/a^{2}$, whose specific
weight decreases as the scale $a$ expressed in  units of $l_{min}$
is going up.

\section{QMFL as a deformation of QM}
\subsection{Main Definitions}
Here we describe QMFL as a deformation of QM using the
above-developed formalism of density pro-matrix. Within it density
pro-matrix should be understood as a deformed density matrix in
QMFL. As fundamental parameter of deformation we use the quantity
$\alpha=l_{min}^{2 }/x^{2}$, where $x$ is the scale. The following
deformation is not claimed as the only one satisfying all the
above properties. Of course, some other deformations are also
possible. At the same time, it seems most natural in a sense that
it allows for minimum modifications of the conventional density
matrix in QM.

\noindent {\bf Definition 1.}\cite{shalyt1}-\cite{shalyt5}

\noindent Any system in QMFL is described by a density pro-matrix
of the form $\rho(\alpha)=\sum_{i}\omega_{i}(\alpha)|i><i|$, where
\begin{enumerate}
\item $0<\alpha\leq1/4$.
\item Vectors $|i>$ form a full orthonormal system;
\item Coefficients $\omega_{i}(\alpha)\geq 0$ and for all $i$  the
 limit $\lim\limits_{\alpha\rightarrow
0}\omega_{i}(\alpha)=\omega_{i}$ exists;
\item
$Sp[\rho(\alpha)]=\sum_{i}\omega_{i}(\alpha)<1$,
$\sum_{i}\omega_{i}=1$.
\item For every operator $B$ and any $\alpha$ there is a
mean operator $B$ depending on  $\alpha$:\\
$$<B>_{\alpha}=\sum_{i}\omega_{i}(\alpha)<i|B|i>.$$
\end{enumerate}
Finally, in order that our definition 1 be in agreement with the
result of section 2, the following condition must be fulfilled:
\begin{equation}\label{U13}
Sp[\rho(\alpha)]-Sp^{2}[\rho(\alpha)]\approx\alpha.
\end{equation}
Hence we can find the value for $Sp[\rho(\alpha)]$ satisfying the
condition of definition 1:
\begin{equation}\label{U14}
Sp[\rho(\alpha)]\approx\frac{1}{2}+\sqrt{\frac{1}{4}-\alpha}.
\end{equation}

According to statement 5  $<1>_{\alpha}=Sp[\rho(\alpha)]$.
Therefore, for any scalar quantity $f$ we have $<f>_{\alpha}=f
Sp[\rho(\alpha)]$. In particular, the mean value
$<[x_{\mu},p_{\nu}]>_{\alpha}$ is equal to
\begin{equation}\label{U15sup}
<[x_{\mu},p_{\nu}]>_{\alpha}= i\hbar\delta_{\mu,\nu}
Sp[\rho(\alpha)].
\end{equation}
We denote the limit $\lim\limits_{\alpha\rightarrow
0}\rho(\alpha)=\rho$ as the density matrix. Evidently, in the
limit $\alpha\rightarrow 0$ we return to QM.

As follows from definition 1,
$<|j><j|>_{\alpha}=\omega_{j}(\alpha)$, from whence the
completeness condition by $\alpha$ is
\\$<(\sum_{i}|i><i|)>_{\alpha}=<1>_{\alpha}=Sp[\rho(\alpha)]$. The
norm of any vector $|\psi>$ assigned to  $\alpha$ can be defined
as

$$<\psi|\psi>_{\alpha}=<\psi|(\sum_{i}|i><i|)_{\alpha}|\psi>
=<\psi|(1)_{\alpha}|\psi>=<\psi|\psi> Sp[\rho(\alpha)],$$

 where
$<\psi|\psi>$ is the norm in QM, i.e. for $\alpha\rightarrow 0$.
Similarly, the described theory may be interpreted using a
probabilistic approach, however requiring  replacement of $\rho$
by $\rho(\alpha)$ in all formulae.

\renewcommand{\theenumi}{\Roman{enumi}}
\renewcommand{\labelenumi}{\theenumi.}
\renewcommand{\labelenumii}{\theenumii.}

\subsection{Some obvious implications}
It should be noted:

\begin{enumerate}
\item The above limit covers both Quantum
and Classical Mechanics. Indeed, since $\alpha\sim L_{p}^{2 }/x^{2
}=G \hbar/c^3 x^{2 }$, we obtain:
\begin{enumerate}
\item $(\hbar \neq 0,x\rightarrow
\infty)\Rightarrow(\alpha\rightarrow 0)$ for QM;
\item $(\hbar\rightarrow 0,x\rightarrow
\infty)\Rightarrow(\alpha\rightarrow 0)$ for Classical Mechanics;
\end{enumerate}
\item As a matter of fact, the deformation parameter $\alpha$
should assume the value $0<\alpha\leq1$.  However, as seen from
(\ref{U14}), $Sp[\rho(\alpha)]$ is well defined only for
$0<\alpha\leq1/4$.That is if $x=il_{min}$ and $i\geq 2$, then
there is not any problem. At the point where $x=l_{min}$ there is
a singularity related to complex values assumed by
$Sp[\rho(\alpha)]$ , i.e. to the impossibility of obtaining a
diagonalized density pro-matrix at this point over the field of
real numbers. For this reason definition 1 has no sense at the
point $x=l_{min}$.We will return to this question when considering
singularity and hypothesis of cosmic censorship in the following
section.
\item We consider possible solutions for (\ref{U13}).
For instance, one of the solutions of (\ref{U13}), at least to the
first order in $\alpha$, is $$\rho^{*}(\alpha)=\sum_{i}\alpha_{i}
exp(-\alpha)|i><i|,$$ where all $\alpha_{i}>0$ are independent of
$\alpha$  and their sum is equal to 1. In this way
$Sp[\rho^{*}(\alpha)]=exp(-\alpha)$. Indeed, we can easily verify
that \begin{equation}\label{U15}
Sp[\rho^{*}(\alpha)]-Sp^{2}[\rho^{*}(\alpha)]=\alpha+O(\alpha^{2}).
\end{equation}
The exponential ansatz for $\rho^{*}(\alpha)$ given here will be
included in subsequent sections. Note that in the momentum
representation $\alpha=p^{2}/p_{max}^{2}\sim p^{2}/p^{2}_{pl}$,
where $p_{pl}$ is the Planck's momentum. When present in matrix
elements, $exp(-\alpha)$ can damp the contribution of great
momenta in a perturbation theory.
\item It is clear that within the proposed description the
states with a unit probability, i.e. pure states, can appear only
in the limit $\alpha\rightarrow 0$, when all $\omega_{i}(\alpha)$
except one are equal to zero or when they tend to zero at this
limit. In our treatment pure states are states, which can be
represented in the form $|\psi><\psi|$, where $<\psi|\psi>=1$.

\item We suppose that all definitions concerning a
density matrix can be carried over to the above-mentioned
deformation of Quantum Mechanics (QMFL)  changing the density
matrix $\rho$ by the density pro-matrix $\rho(\alpha)$ with
subsequent passing to the low-energy limit $\alpha\rightarrow 0$.
Specifically, for statistical entropy we have
\begin{equation}\label{U16}
S_{\alpha}=-Sp[\rho(\alpha\ln(\rho(\alpha))].
\end{equation}
The quantity of $S_{\alpha}$ seems never to be equal to zero as
$\ln(\rho(\alpha))\neq 0$ and hence $S_{\alpha}$ may be equal to
zero at the limit $\alpha\rightarrow 0$ only.
\end{enumerate}
The following statements are essential for our study:
\begin{enumerate}
\item If we carry out a measurement at a pre-determined scale, it is
impossible to regard the density pro-matrix as a density matrix
with an accuracy better than the limit $\sim10^{-66+2n}$, where
$10^{-n}$ is the measuring scale. In the majority of known cases
this is sufficient to consider the density pro-matrix as a density
matrix. But at Planck's scale, where quantum gravitational effects
and Planck's energy levels cannot be neglected, the difference
between $\rho(\alpha)$ and  $\rho$ should be taken into
consideration.

\item Proceeding from the above, on Planck's scale the
notion of Wave Function of the Universe (as introduced in
\cite{r11}) has no sense, and quantum gravitation effects in this
case should be described with the help of density pro-matrix
$\rho(\alpha)$ only.
\item Since density pro-matrix $\rho(\alpha)$ depends on the measuring
scale, evolution of the Universe within the inflationary model
paradigm \cite{r10} is not a unitary process, or otherwise the
probabilities $p_{i}=\omega_{i}(\alpha)$  would be preserved.
\end{enumerate}

\section{Applications of the Quantum-Mechanical Density Pro-Matrix}
In this section some apparent applications of the primary
definitions and methods derived in the previous section are given
\cite{shalyt3}-\cite{shalyt5}.
\subsection{Dynamic aspects of QMFL. Deformed Liouville  equation}
 Let's suppose that in QMFL a
density pro-matrix has the form $\rho[\alpha(t),t]$, in other
words, it depends on two parameters: time $t$ and parameter of
deformation $\alpha$, which also depends on time
($\alpha=\alpha(t)$). Then, we have
\begin{equation}\label{U17}
\rho[\alpha(t),t]=\sum\omega_{i}[\alpha(t)]|i(t)><i(t)|.
\end{equation}
Differentiating the last expression with respect to time, we
obtain
\begin{equation}\label{U18}
\frac{d\rho}{dt}=\sum_{i}
\frac{d\omega_{i}[\alpha(t)]}{dt}|i(t)><i(t)|-i[H,\rho(\alpha)]=d[ln\omega(\alpha)]\rho
(\alpha)-i[H,\rho(\alpha)].
\end{equation}
Where $ln[\omega(\alpha)]$ is a row-matrix and $\rho(\alpha)$ is a
column-matrix. In such a way we have obtained a prototype of
 Liouville's equation.

Let's consider some  cases of particular importance.
\begin{enumerate}
\item First we consider the process of time
evolution at low energies, i.e. when $\alpha \approx 0$,
$\alpha(t)\approx 0$ and $t \to \infty$. Then it is clear that
$\omega_{i}(\alpha)\approx \omega_{i} \approx constant$. The first
term in (\ref{U18}) vanishes and we obtain  Liouville equation.
\item Also we obtain  the Liouville's equation when using
inflationary approach and going to large-scales. Within the
inflationary approach the scale $a \approx e^{Ht}$, where $H$ is
the Hubble's constant and $t$ is time. Therefore $\alpha \sim
e^{-2Ht}$ and when $t$ is quite big $\alpha\to 0$. In other words,
$\omega_{i}(\alpha) \to \omega_{i}$, the first term in (\ref{U18})
vanishes and again  we obtain the Liouville's equation.
\item At very early stage of the inflationary process  or even before it
took place $\omega_{i}(\alpha)$ was not a constant and hense, the
first term in (\ref{U18}) should be taking into account. This way
we obtain a deviation from the Liouville's equation.
\item Finally, let us consider the case when $\alpha(0) \approx 0$,
$\alpha(t)>0$ where $t \to \infty$. In this case we are going from
low-energy to high-energy scale one and $\alpha(t)$ grows when $t
\to \infty$. The first term in (\ref{U18}) is significant and we
obtain an addition to the Liouville's equation of the form
$$d[ln\omega(\alpha)]\rho(\alpha).$$ This could be the case  when
matter goes into a black hole and is moving in direction of the
singularity (to the Planck's scale).
\end{enumerate}

 \subsection{Singularity, entropy and information loss in black
holes}
Note that remark II in section 3.2 about complex meaning
assumed by the density pro-matrix at the point with fundamental
length is directly related to the singularity problem and cosmic
censorship in the General Theory of Relativity \cite{r12}.  For
instance, considering a Schwarzchild's black hole (\cite{r13})
with metrics:
\begin{equation}\label{U19}
 ds^2 = - (1 - \frac{2M}{r}) dt^2 +
\frac{dr^2}{(1 - \frac{2M}{r})} + r^2 d \Omega_{II}^2,
\end{equation}
we obtain a well-known a singularity at the point $r=0$. In our
approach this corresponds to the point with fundamental length
($r=l_{min}$). At this point we are not able to measure anything,
since at this point $\alpha=1$ and $Sp[\rho (\alpha)]$ becomes
complex. Thus, we carry out a measurement, starting from the point
$r=2l_{min}$ that corresponds to $\alpha=1/4$. Consequently, the
initial singularity $r=l_{min}$, which cannot be measured, is
hidden of observation. This could confirm the hypothesis of cosmic
censorship in this particular  case. By this hypothesis  "a naked
singularity cannot be observed". Thus, QMFL in our approach
"feels" the singularity  compared with QM, that does not
\cite{shalyt4,shalyt5}. Statistical entropy, associated with the
density pro-matrix and introduced in the remark V section 3 is
written $$S_{\alpha}=-Sp[\rho(\alpha)\ln(\rho(\alpha))],$$ and may
be interpreted as a density of entropy on the unit  minimal area
$l^{2}_{min}$ depending on the scale $x$. It could be quite big
close to the singularity, i.e. for $\alpha\rightarrow 1/4$. This
does not contradict the second law of Thermodynamics since maximal
entropy of a specific  object in the Universe is proportional to
the square of its surface $A$, measured in units of minimal square
$l^{2}_{min}$ or Planck's square $L_{p}^2$, as  shown in some
papers (see, for instance \cite{r14}). Therefore, in the expanded
Universe since surface $A$ grows, entropy does not decrease.
\\The obtained results enable one to consider anew the information loss
problem associated with black holes \cite{r15,r16}, at least, for
the case of "mini" black holes \cite{shalyt4,shalyt5}. Indeed,
when we consider these black holes, Planck's scale is a factor. It
was shown that entropy of matter absorbed by a black hole at this
scale is not equal to zero, supporting the data of R.Myers
\cite{Myers}. According to his results, the pure state cannot form
a black hole. Then,  it is necessary to reformulate the problem
per se, since so far in all publications on information paradox
zero entropy at the initial state has been compared to  nonzero
entropy at the final state. According to our analysis at the
Planck's scale there is not initial zero entropy and "mini" black
holes with masses of the order $M_{pl}$ should not radiate at all.
Similar results  were deduced by A.D.Helfer\cite{r31} using
another approach: "p.1...The possibility that non-radiating "mini"
black holes should be taken seriously; such holes could be part of
the dark matter in the Universe". Note that in \cite{r31} the main
argument in favor of the existence of non-radiating "mini" black
holes developed with consideration of quantum gravity effects. In
our analysis these effects are considered implicitly since,as
stated above, any approach in quantum gravity leads to the
fundamental-length concept \cite{r1}. Besides, it should be noted
that in some recent papers for all types of black holes QMFL with
GUR is considered from the start \cite{r18},\cite{r32}. By this
approach stable remnants with masses of the order of Planck's ones
$M_{pl}$ emerge during the process of black hole evaporation. From
here it follows that black holes should not evaporate fully. We
arrive to the conclusion that results given in \cite{r13,r19} are
correct only in the semi-classical approximation and they should
not be applicable to the quantum back hole analysis.
\\At least at a qualitative level the above results can clear up
the answer to the question, how information may be lost at big
black holes formed due to the star collapse. Our point of view is
close to that of R.Penrose's one \cite{r20} who considers  that
information in black holes is lost when matter meets a
singularity. In our approach information loss takes place in the
same form. Indeed, near  the horizon of events an approximately
pure state with the initial entropy practically equal to zero
$S^{in}=-Sp[\rho\ln(\rho)]$, that corresponds to $\alpha \to 0$,
when approaching a singularity (of reaching the Planck's scale)
gives yet non zero entropy
$S_{\alpha}=-Sp[\rho(\alpha)\ln(\rho(\alpha))]>0$ for $\alpha
>0$. Therefore, entropy increases and information is lost in this
black hole. We can (at the moment, at a qualitative level)
evaluate the entropy of black holes. Actually, starting from a
density matrix for the pure state at the "entry" to a black hole
$\rho_{in}=\rho_{pure}$ with zero entropy $S^{in}=0$, we obtain
with a straightforward "naive" calculation (that is (\ref{U13}) is
considered  an exact relation). Then,for the singularity in the
black hole the corresponding entropy of the density pro-matrix
$Sp[\rho_{out}]=1/2$ at $\alpha=1/4$ is $$S^{out}=S_{1/4}=-1/2
\ln1/2 \approx 0.34657.$$
 Taking into account that total entropy of a
black hole is proportional to the quantum area of surface A,
measured in Planck's units of area $L_{p}^2$ \cite{r21}, we obtain
the following value for quantum entropy of a black hole:
\begin{equation}\label{U20}
S_{BH}= 0.34657 \frac{A}{L_{p}^2}
\end{equation}

This formula differs from the well-known one given by
Bekenstein-Hawking for black hole entropy $S_{BH}=\frac{1}{4}
\frac{A}{L_{p}^2}$ \cite{r22}. This result was obtained in the
semi-classical approximation. At the present moment quantum
corrections to this formula are an object of investigation
\cite{r23}. As it was mentioned above we carry out straightforward
calculation. Otherwise, using the ansatz of statement  remark III
in section 3 and assuming that spur of density pro-matrix is equal
to $Sp[\rho^{*}(\alpha)]=exp(-\alpha)$, we obtain for $\alpha=1/4$
that entropy is equal to $$S^{* out}=S^{*}_{1/4}=-Sp[exp(-1/4)\ln
exp(-1/4)]\approx 0.1947,$$ and consequently we arrive to the
following value of entropy
\begin{equation}\label{U21}
S_{BH} = 0.1947 \frac{A}{L_{p}^2}
\end{equation}
that is the closest  the result obtained in \cite{r23}. Our
approach leading to formula (\ref{U21}) is from the very beginning
based on quantum nature of black holes. Note here that in the
approaches used up to now to modify Liouville's equation due to
information paradox \cite{r24} the added member appearing in the
right side of (\ref{U18}) takes the form $$-\frac{1}{2}\sum_{\xi
\gamma \neq 0} (Q^{\gamma}Q^{\xi}\rho+\rho Q^{\gamma}Q^{\xi}-2
Q^{\xi}\rho Q^{\gamma}),$$ where $Q^{\xi}$
  is a full orthogonal set
of Hermitian matrices with $Q^{0} =1$. In this case either
locality or conservation of energy-impulse tensor is broken down.
By the approach offered in this paper the member added in the
deformed Liouville's equation,in our opinion, has a more natural
and beautiful form: $$d[ln\omega(\alpha)]\rho (\alpha).$$ In the
limit $\alpha\to 0$ all properties of QM are conserved, the added
member vanishes and we obtain Liouville's equation.
\\The information paradox problem at black holes is considered in
greater detail in section 7, where the above methods provide a new
approach to  this problem.

\subsection{Bekenstein-Hawking formula}
The problem is whether can the well-known semiclassical
Bekenstein-Hawking formula for Black Hole  entropy
\cite{r21},\cite{r32} can be obtained within the proposed approach
? We show how  to do it \cite{shalyt5}. To obtain black hole
quantum entropy, we use the formula
  $S_{\alpha}=-Sp[\rho(\alpha)\ln(\rho(\alpha))]=-<\ln(\rho(\alpha))>_{\alpha}$
    when $\alpha$ takes its maximal meaning ($\alpha = 1/4$).
   In this case (\ref{U20}) and (\ref{U21}) can be written as
\begin{equation}\label{U22}
S_{BH} = -<\ln(\rho(1/4))>_{1/4} \frac{A}{L_{p}^2},
\end{equation}
for different $\rho(\alpha)$ in  (\ref{U20}) and  (\ref{U21}) but
for the same value of $\alpha$ ($\alpha = 1/4$). Semiclassical
approximation works only at large-scales, therefore measuring
procedure is also defined at large scales. In other words, all
mean values must be taken when $\alpha = 0$. However, for the
operators whose mean values are calculated the dependence on
$\alpha$ should be taken  into account since according to the
well-known Hawking's paper \cite{r14}, operator of superscattering
$\$$ translates $\$:\rho_{in}\mapsto\rho_{out}$, where in the case
considered $\rho_{in}=\rho_{pure}$ and
$\rho_{out}=\rho_{pure}^{*}(\alpha)=
exp(-\alpha)\rho_{pure}=exp(-1/4)\rho_{pure}$ conforming to the
exponential ansatz of statement  III, section 3. Therefore we have
\\
$$S^{semiclass}_{\alpha}=-<\ln(\rho(\alpha))>$$
\\
and formula for semiclassical entropy of a black hole takes the
form
\begin{equation}\label{U23}
S^{semiclass}_{BH} = -<\ln(\rho(1/4))>
\frac{A}{L_{p}^2}=-<ln[exp(-1/4)]\rho_{pure}>\frac{A}{L_{p}^2}
=\frac{1}{4}\frac{A}{L_{p}^2}
\end{equation}
that coincides with the well-known Bekenstein-Hawking formula. It
should be noted that  $\alpha = 1/4$  in our approach appears in
section 3 quite naturally as a maximal meaning for which
$Sp\rho(\alpha)$ still stays real, according to (\ref{U13}) and
(\ref{U14}). Apparently, if considering corrections of order
higher than 1 on $\alpha$, then members from $O(\alpha^{2})$ in
the formula for $\rho_{out}$ in (\ref{U15}) can give quantum
corrections \cite{r23} for $S^{semiclass}_{BH}$ (\ref{U23}) in our
approach.

\subsection{Some comments on Shr{\"o}dinger's picture}
 As it was indicated above in the
statement 1 section 3.2, we are able to obtain from QMFL two
limits: Quantum and Classical Mechanics. The deformation described
here should be understood as "minimal" in the sense that we have
deformed only the probability $\omega_{i}\rightarrow
\omega_{i}(\alpha)$, whereas the state vectors have been not
deformed. In a most complete treatment we have to consider vectors
$|i(\alpha)><i(\alpha)|$ instead $|i><i|$,  and in this case the
full picture will be very complicated. It is easy to understand
how Shrodinger's picture is transformed in QMFL \cite{shalyt5}.
The prototype of Quantum Mechanical normed wave function $\psi(q)$
with $\int|\psi(q)|^{2}dq=1$ in QMFL is $\theta(\alpha)\psi(q)$.
The deformation parameter   $\alpha$ assumes the value
$0<\alpha\leq1/4$. Its properties are
$|\theta(\alpha)|^{2}<1$,$\lim\limits_{\alpha\rightarrow
0}|\theta(\alpha)|^{2}=1$ and the relation
$|\theta(\alpha)|^{2}-|\theta(\alpha)|^{4}\approx \alpha$ takes
place. In such a way the full probability always is less than 1:
$p(\alpha)=|\theta(\alpha)|^{2}=\int|\theta(\alpha|^{2}|\psi(q)|^{2}dq<1$
tending to 1 when  $\alpha\rightarrow 0$. In the most general case
of arbitrarily normed state in QMFL
$\psi=\psi(\alpha,q)=\sum_{n}a_{n}\theta_{n}(\alpha)\psi_{n}(q)$
with $\sum_{n}|a_{n}|^{2}=1$ the full probability is
$p(\alpha)=\sum_{n}|a_{n}|^{2}|\theta_{n}(\alpha)|^{2}<1$ and
 $\lim\limits_{\alpha\rightarrow 0}p(\alpha)=1$.

 It is natural that in QMFL Shrodinger's equation is also
deformed. It is replaced by  equation
\begin{equation}\label{U24}
\frac{\partial\psi(\alpha,q)}{\partial t}
=\frac{\partial[\theta(\alpha)\psi(q)]}{\partial
t}=\frac{\partial\theta(\alpha)}{\partial
t}\psi(q)+\theta(\alpha)\frac{\partial\psi(q)}{\partial t},
\end{equation}
where the second term in the right side generates the Shrodinger's
equation since
\begin{equation}\label{U25}
\theta(\alpha)\frac{\partial\psi(q)}{\partial
t}=\frac{-i\theta(\alpha)}{\hbar}H\psi(q).
\end{equation}

Here $H$ is the Hamiltonian and the first member is added,
similarly to the member appearing in the deformed Loiuville's
equation and  vanishing when $\theta[\alpha(t)]\approx const$. In
particular, this takes place in the low energy limit in QM, when
$\alpha\rightarrow 0$.  Note that the above theory  is not a
time-reversal as QM, since the combination $\theta(\alpha)\psi(q)$
breaks down this property in the deformed Shrodinger's equation.
Time-reversal is conserved only in the low energy limit, when
quantum mechanical Shrodinger's equation is valid.

\section{Density Matrix Deformation in Statistical Mechanics of
Early Universe}
\subsection{Main definition and properties}
First we revert to  the Generalized Uncertainty Relations
 "coordinate - momentum" (section 2,formula (\ref{U2})) :
\begin{equation}\label{U5s}
\triangle x\geq\frac{\hbar}{\triangle p}+\alpha^{\prime}
L_{p}^2\frac{\triangle p}{\hbar}.
\end{equation}
 Using relations (\ref{U5s}) it is easy to obtain a similar relation for the
 "energy - time" pair. Indeed (\ref{U5s}) gives
\begin{equation}\label{U6s}
\frac{\Delta x}{c}\geq\frac{\hbar}{\Delta p c }+\alpha^{\prime}
L_{p}^2\,\frac{\Delta p}{c \hbar},
\end{equation}
then
\begin{equation}\label{U7s}
\Delta t\geq\frac{\hbar}{\Delta
E}+\alpha^{\prime}\frac{L_{p}^2}{c^2}\,\frac{\Delta p
c}{\hbar}=\frac{\hbar}{\Delta E}+\alpha^{\prime}
t_{p}^2\,\frac{\Delta E}{ \hbar}.
\end{equation}
where the smallness of $L_p$ is taken into account so that the
difference between $\Delta E$ and $\Delta (pc)$ can be neglected
and $t_{p}$  is the Planck time
$t_{p}=L_p/c=\sqrt{G\hbar/c^5}\simeq 0,54\;10^{-43}sec$. From
whence it follows that we have a  maximum energy of the order of
Planck's:
\\
$$E_{max}\sim E_{p}$$
\\
Proceeding to the Statistical Mechanics, we further assume that an
internal energy of any ensemble U could not be in excess of
$E_{max}$ and hence temperature $T$ could not be in excess of
$T_{max}=E_{max}/k_{B}\sim T_{p}$. Let us consider density matrix
in Statistical Mechanics (see \cite{r34}, Section 2, Paragraph 3):
\begin{equation}\label{U8s}
\rho_{stat}=\sum_{n}\omega_{n}|\varphi_{n}><\varphi_{n}|,
\end{equation}
where the probabilities are given by
\\
$$\omega_{n}=\frac{1}{Q}\exp(-\beta E_{n})$$ and
\\
$$Q=\sum_{n}\exp(-\beta E_{n}).$$
\\
Then for a canonical Gibbs ensemble the value
\begin{equation}\label{U9s}
\overline{\Delta(1/T)^{2}}=Sp[\rho_{stat}(\frac{1}{T})^{2}]
-Sp^{2}[\rho_{stat}(\frac{1}{T})],
\end{equation}
is always equal to zero, and this follows from the fact that
$Sp[\rho_{stat}]=1$. However, for very high temperatures $T\gg0$
we have $\Delta (1/T)^{2}\approx 1/T^{2}\geq 1/T_{max}^{2}$. Thus,
for $T\gg0$ a statistical density matrix $\rho_{stat}$ should be
deformed so that in the general case \cite{shalyt6,shalyt7}
\begin{equation}\label{U10s}
Sp[\rho_{stat}(\frac{1}{T})^{2}]-Sp^{2}[\rho_{stat}(\frac{1}{T})]
\approx \frac{1}{T_{max}^{2}},
\end{equation}
or \begin{equation}\label{U11s}
Sp[\rho_{stat}]-Sp^{2}[\rho_{stat}] \approx
\frac{T^{2}}{T_{max}^{2}}.
\end{equation}
In this way $\rho_{stat}$ at very high $T\gg 0$ becomes dependent
on the parameter $\tau = T^{2}/T_{max}^{2}$, i.e. in the most
general case
\\
$$\rho_{stat}=\rho_{stat}(\tau)$$ and $$Sp[\rho_{stat}(\tau)]<1$$
\\
and for $\tau\ll 1$ we have $\rho_{stat}(\tau)\approx\rho_{stat}$
(formula (\ref{U8s})) .\\ This situation is identical to the case
associated with the deformation parameter $\alpha = l_{min}^{2
}/x^{2}$ of QMFL given in section â 3. That is the condition
$Sp[\rho_{stat}(\tau)]<1$ has an apparent physical meaning when:
\begin{enumerate}
 \item At temperatures close to $T_{max}$ some portion of information
about the ensemble is inaccessible in accordance with the
probability that is less than unity, i.e. incomplete probability.
 \item And vice versa, the longer is the distance from $T_{max}$ (i.e.
when approximating the usual temperatures), the greater is the
bulk of information and the closer is the complete probability to
unity.
\end{enumerate}
 Therefore similar to the introduction of the deformed
quantum-mechanics density matrix in section 3 we give the
following
\\
\noindent {\bf Definition 2.} {\bf(Deformation of Statistical
Mechanics)}\cite{shalyt6,shalyt7,shalyt8} \noindent \\Deformation
of Gibbs distribution valid for temperatures on the order of the
Planck's $T_{p}$ is described
 by deformation of a statistical density matrix
  (statistical density pro-matrix) of the form
\\$${\bf \rho_{stat}(\tau)=\sum_{n}\omega_{n}(\tau)|\varphi_{n}><\varphi_{n}|}$$
 having the deformation parameter
$\tau = T^{2}/T_{max}^{2}$, where
\begin{enumerate}
\item $0<\tau \leq 1/4$.
\item The vectors $|\varphi_{n}>$ form a full orthonormal system;
\item $\omega_{n}(\tau)\geq 0$ and for all $n$ at $\tau \ll 1$
 we obtain
 $\omega_{n}(\tau)\approx \omega_{n}=\frac{1}{Q}\exp(-\beta E_{n})$
In particular, $\lim\limits_{T_{max}\rightarrow \infty
(\tau\rightarrow 0)}\omega_{n}(\tau)=\omega_{n}$
\item
$Sp[\rho_{stat}]=\sum_{n}\omega_{n}(\tau)<1$,
$\sum_{n}\omega_{n}=1$;
\item For every operator $B$ and any $\tau$ there is a
mean operator $B$ depending on  $\tau$ \\
$$<B>_{\tau}=\sum_{n}\omega_{n}(\tau)<n|B|n>.$$
\end{enumerate}
Finally, in order that our Definition 2 agree with the formula
(\ref{U11s}), the following condition must be fulfilled:
\begin{equation}\label{U12s}
Sp[\rho_{stat}(\tau)]-Sp^{2}[\rho_{stat}(\tau)]\approx \tau.
\end{equation}
Hence we can find the value for $Sp[\rho_{stat}(\tau)]$
 satisfying
the condition of Definition 2 (similar to Definition 1):
\begin{equation}\label{U13s}
Sp[\rho_{stat}(\tau)]\approx\frac{1}{2}+\sqrt{\frac{1}{4}-\tau}.
\end{equation}
It should be noted:

\begin{enumerate}
\item The condition $\tau \ll 1$ means that $T\ll T_{max}$ either
$T_{max}=\infty$ or both in accordance with a normal Statistical
Mechanics and canonical Gibbs distribution (\ref{U8s})
\item Similar to QMFL in Definition 1, where the deformation
parameter $\alpha$ should assume the value $0<\alpha\leq1/4$. As
seen from (\ref{U13s}), here $Sp[\rho_{stat}(\tau)]$ is well
defined only for $0<\tau\leq1/4$. This means that the feature
occurring in QMFL at the point of the fundamental length
$x=l_{min}$ in the case under consideration is associated with the
fact that {\bf highest  measurable temperature of the ensemble is
always} ${\bf T\leq \frac{1}{2}T_{max}}$.

\item The constructed deformation contains all four fundamental constants:
 $G,\hbar,c,k_{B}$ as $T_{max}=\varsigma T_{p}$,where $\varsigma$
 is the denumerable function of  $\alpha^{\prime}$
(\ref{U5s})and $T_{p}$, in its turn, contains all the
above-mentioned
 constants.

\item Again similar to QMFL, as a possible solution for (\ref{U12s})
we have an exponential ansatz
\\
$$\rho_{stat}^{*}(\tau)=\sum_{n}\omega_{n}(\tau)|n><n|=\sum_{n}
exp(-\tau) \omega_{n}|n><n|$$
\\
\begin{equation}\label{U14s}
Sp[\rho_{stat}^{*}(\tau)]-Sp^{2}[\rho_{stat}^{*}(\tau)]=\tau+O(\tau^{2}).
\end{equation}
In such a way with the use of an exponential ansatz (\ref{U14s})
the deformation of a canonical Gibbs distribution at Planck scale
(up to factor $1/Q$) takes an elegant and completed form:
\begin{equation}\label{U15s}
{\bf \omega_{n}(\tau)=exp(-\tau)\omega_{n}= exp(-\frac{T^{2}}
{T_{max}^{2}}-\beta E_{n})}
\end{equation}
where $T_{max}= \varsigma T_{p}$
\end{enumerate}
\subsection{Some implications}
Using in this section only the
exponential ansatz of (\ref{U14s}), in the coordinate
representation we have the following:
\begin{equation}\label{U16s}
\rho(x,x^{\prime},\tau)=\sum_{i}\frac{1}{Q}e^{-\beta
E_{i}-\tau}\varphi_{i}(x)\varphi_{i}^{*}(x^{\prime})
\end{equation}
However, as $H \mid \varphi_{i}>=E_{i} \mid \varphi_{i}>$, then
\begin{equation}\label{U17s}
\rho(\beta,\tau)=\frac{1}{Q}\sum_{i}e^{-\beta H-\tau}\mid
\varphi_{i}><\varphi_{i}\mid=\frac{e^{-\beta H-\tau}}{Q},
\end{equation}
where $Q=\sum_{i}e^{-\beta E_{i}}=Spe^{-\beta H}$. Consequently,
\begin{equation}\label{U18s}
\rho(\beta,\tau)=\frac{e^{-\beta H-\tau}}{Spe^{-\beta H}}
\end{equation}
In this way the "deformed" average energy of a system is obtained
as
\begin{equation}\label{U19s}
U_{\tau}=Sp\rho(\tau)H=\frac{He^{-\beta H-\tau}}{Spe^{-\beta H}}
\end{equation}
The calculation of "deformed" entropy is also a simple task.
Indeed, in the general case of the canonical Gibbs distribution
the probabilities are equal to
\begin{equation}\label{U20s}
P_{n}=\frac{1}{Q}e^{-\beta E_{n}}
\end{equation}
Nevertheless, in case under consideration they are "replenished"
by $exp(-\tau)$ factor and hence are equal to
\begin{equation}\label{U21s}
P^{\tau}_{n}=\frac{1}{Q}e^{-(\tau+\beta E_{n})}.
\end{equation}
Thus, a new formula for entropy in this case is as follows:
\begin{equation}\label{U22s}
S_{\tau}=-k_{B}e^{-\tau}\sum_{n}P_{n}(lnP_{n}-\tau)
\end{equation}
It is obvious that
 $\lim\limits_{\tau\rightarrow 0}S_{\tau}=
S$, where $S$ - entropy of the canonical ensemble, that is a
complete analog of its counterpart in quantum mechanics at the
Planck scale $\lim\limits_{\alpha\rightarrow 0}S_{\alpha}= S$,
where $S$ - statistical entropy in quantum mechanics, and
deformation parameter $\tau$ is changed by $\alpha$ of section 3.
\\Given the average energy deformation in a system $U_{\tau}$ and
knowing the entropy deformation, one is enabled to calculate the
"deformed" free energy $F_{\tau}$ as well:
\begin{equation}\label{U23s}
F_{\tau}=U_{\tau}-TS_{\tau}
\end{equation}
Consider the counterpart of Liouville equation \cite{r34} for the
unnormed $\rho(\beta,\tau)$ (\ref{U18s}):
\begin{equation}\label{U24s}
-\frac{\partial\rho(\beta,\tau)}{\partial\beta}=
-\frac{\partial}{\partial\beta}e^{-\tau-\beta H},
\end{equation}
where
\\
$$\tau=\frac{T^{2}}{T_{max}^{2}}=\frac{\beta^{2}_{max}}{\beta^{2}},$$
\\
where $\beta_{max}=1/k_{B}T_{max}\sim 1/k_{B}T_{P}\equiv
\beta_{P}$, $\tau=\tau(\beta)$. Taking this into consideration and
expanding the right-hand side of equation (\ref{U24s}), we get
deformation  of Liouville equation further referred to as
$\tau$-deformation:
\begin{equation}\label{U25s}
-\frac{\partial\rho(\beta,\tau)}{\partial\beta}=
-e^{-\tau}\frac{\partial \tau}{\partial \beta}+e^{-\tau}H
\rho(\beta)=e^{-\tau}[H \rho(\beta)-\frac{\partial \tau}{\partial
\beta}],
\end{equation}
where $\rho(\beta)=\rho(\beta,\tau=0)$.
\\ The first term in brackets (\ref{U25s}) generates Liouville equation.
Actually, taking the limit of the left and right sides
(\ref{U25s}) for $\tau\rightarrow 0$, we derive the normal
Liouville equation for $\rho(\beta)$ in statistical mechanics
\cite{r34}:
\begin{equation}\label{U26s}
-\frac{\partial\rho(\beta)}{\partial\beta}=H \rho(\beta)
\end{equation}
By this means we obtain a complete analog of the
quantum-mechanical results for the associated deformation of
Liouville equation derived  in section 4.1 and
\cite{shalyt3}-\cite{shalyt5}.
 Namely:
\\
(1)Early Universe (scales approximating those of the Planck's,
original singularity, $\tau>0$). The density pro-matrix
$\rho(\beta,\tau)$ is introduced and a $\tau$-deformed Liouville
equation(\ref{U25s}), respectively;
\\
(2)after the inflation extension (well-known scales, $\tau\approx
0$) the normal density matrix $\rho(\beta)$ appears in the limit
$\lim\limits_{\tau\rightarrow 0}\rho(\beta,\tau)=\rho(\beta)$.
$\tau$- deformation of Liouville equation (\ref{U25s}) is changed
by a well-known Liouville equation(\ref{U26s});
\\
(3)and finally the case of the matter absorbed by a black hole and
its tendency to the singularity. Close to the black hole
singularity both quantum and statistical mechanics are subjected
to deformation as they do in case of the original singularity
\cite{shalyt3}-\cite{shalyt5}. Introduction of temperature on the
order of the Planck's \cite{Castro1},\cite{Castro2} and hence the
deformation parameter $\tau > 0$ may be taken as an indirect
evidence for the fact. Because of this, the case is associated
with the reverse transition from the well-known density matrix in
statistical mechanics $\rho(\beta)$ to its $\tau$-deformation
$\rho(\beta,\tau)$ and from Liouville equation(\ref{U26s}) to its
$\tau$-deformation (\ref{U25s}).

\section{Generalized Uncertainty Relation
\\ in Thermodynamics}

 Now we consider the thermodynamic uncertainty relations between the
inverse temperature and interior energy of a macroscopic ensemble
\begin{equation}\label{U12t}
\Delta \frac{1}{T}\geq\frac{k}{\Delta U},
\end{equation}
where $k$ is the Boltzmann constant. \\ N.Bohr \cite{Bohr1} and
W.Heisenberg \cite{Heis1} first pointed out that such kind of
uncertainty principle should take place in thermodynamics. The
thermodynamic uncertainty  relations (\ref{U12t})  were proved by
many authors and in various ways \cite{Uncert1}. Therefore their
validity does not raise any doubts. Nevertheless, relation
(\ref{U12t}) was established using a standard model for the
infinite-capacity heat bath encompassing the ensemble. But it is
obvious from the above inequalities that at very high energies the
capacity of the heat bath can no longer be assumed infinite at the
Planck scale. Indeed, the total energy of the pair heat bath -
ensemble may be arbitrary large but finite, merely as the Universe
is born at a finite energy. Thus the quantity that can be
interpreted as a temperature of the ensemble must have the upper
limit and so does its main quadratic deviation. In other words the
quantity $\Delta (1/T)$ must be bounded from below. But in this
case an additional term should be introduced into
(\ref{U12t})\cite{shalyt9, shalyt10, shalyt7}
\begin{equation}\label{U12at}
\Delta \frac{1}{T}\geq\frac{k}{\Delta U} + \eta\,\Delta U,
\end{equation}
where $\eta$ is a coefficient. Dimension and symmetry reasons give
$$ \eta \sim \frac{k}{E_p^2}\enskip or\enskip \eta =
\alpha^{\prime} \frac{k}{E_p^2} $$
 As in the previous cases
inequality (\ref{U12at}) leads to the fundamental (inverse)
temperature.
\begin{equation}\label{U15t}
T_{max}=\frac{\hbar}{2\surd \alpha^{\prime}t_{p}
k}=\frac{\hbar}{\Delta t_{min} k}, \quad \beta_{min} = {1\over
kT_{max}} =  \frac{\Delta t_{min}}{\hbar}
\end{equation}
It should be noted that the same conclusion about the existence of
 maximal temperature in Nature can be made also considering black
hole evaporation \cite{Castro3}.
 \\ Thus, we obtain the
system of generalized uncertainty relations in the symmetric form
\begin{equation}\label{U17t}
\left\{
\begin{array}{lll}
\Delta x & \geq & \frac{\displaystyle\hbar}{\displaystyle\Delta
p}+ \alpha^{\prime} \left(\frac{\displaystyle\Delta
p}{\displaystyle P_{pl}}\right)\,
\frac{\displaystyle\hbar}{\displaystyle P_{pl}}+... \\ &  &  \\
\Delta t & \geq & \frac{\displaystyle\hbar}{\displaystyle\Delta
E}+\alpha^{\prime} \left(\frac{\displaystyle\Delta
E}{\displaystyle E_{p}}\right)\,
\frac{\displaystyle\hbar}{\displaystyle E_{p}}+...\\
  &  &  \\
  \Delta \frac{\displaystyle 1}{\displaystyle T}& \geq &
  \frac{\displaystyle k}{\displaystyle\Delta U}+\alpha^{\prime}
  \left(\frac{\displaystyle\Delta U}{\displaystyle E_{p}}\right)\,
  \frac{\displaystyle k}{\displaystyle E_{p}}+...
\end{array} \right.
\end{equation}
or in the equivalent form
\begin{equation}\label{U18t}
\left\{
\begin{array}{lll}
\Delta x & \geq & \frac{\displaystyle\hbar}{\displaystyle\Delta
p}+\alpha^{\prime} L_{p}^2\,\frac{\displaystyle\Delta
p}{\displaystyle\hbar}+... \\
  &  &  \\
  \Delta t & \geq &  \frac{\displaystyle\hbar}{\displaystyle\Delta E}+\alpha^{\prime}
  t_{p}^2\,\frac{\displaystyle\Delta E}{\displaystyle\hbar}+... \\
  &  &  \\

  \Delta \frac{\displaystyle 1}{\displaystyle T} & \geq &
  \frac{\displaystyle k}{\displaystyle\Delta U}+\alpha^{\prime}
  \frac{\displaystyle 1}{\displaystyle T_{p}^2}\,
  \frac{\displaystyle\Delta U}{\displaystyle k}+...
\end{array} \right.
\end{equation}
where dots mean the existence of higher order corrections as in
\cite{r27}.
 Here $T_{p}$ is the Planck temperature:
$T_{p}=\frac{E_{p}}{k}$.
\\In conclusion of this section we would like to note that the restriction on
the heat bath made above makes the equilibrium   partition
function non-Gibbsian \cite{r35}.
\\ Note that the last-mentioned inequality is symmetrical to the second one
with respect to substitution \cite{Castro4}
\\
$$ t\mapsto\frac{1}{T}, \hbar\mapsto k,\triangle E\mapsto
\triangle U . $$ However this observation can by no means be
regarded as a rigorous proof of the generalized uncertainty
relation in thermodynamics.
\\
There is  reason to believe that  rigorous justification for the
latter  (thermodynamic) inequalities in systems (\ref{U17t}) and
(\ref{U18t}) may be made by means of a certain deformation of
Gibbs distribution. One of such deformations that, by the author's
opinion, is liable to give the indicated result has been
considered in the previous section  of this chapter and in some
other papers \cite{shalyt6,shalyt7}.

\section{Non-Unitary and Unitary Transitions
 in \\Generalized Quantum  Mechanics
 and \\Information Problem Solving}
In this section the earlier obtained results are used for the
unitarity study in Generalized Quantum  Mechanics and
 Information Paradox Problem \cite{r15},\cite{r13},\cite{r16}.
 It is demonstrated that the existence of
black holes in the suggested approach in the end twice causes
nonunitary transitions resulting in the unitarity. In parallel
this problem is considered in other terms: entropy density,
Heisenberg algebra deformation terms, respective deformations of
Statistical Mechanics, - all showing the identity of the basic
results. From this an explicit solution for Information Paradox
Problem has been derived. This section is based on the results
presented in \cite{shalyt11,shalyt12,shalyt13}

\subsection{Some comments and unitarity in
QMFL}
 As has been indicated in section 4.4, time reversal
is retained in the large-scale limit only. The same is true for
the superposition principle in Quantum Mechanics. Indeed, it may
be retained in a very narrow interval of cases for the functions
$\psi_{1}(\alpha,q)=\theta(\alpha)\psi_{1}(q)$ è
$\psi_{2}(\alpha,q)=\theta(\alpha)\psi_{2}(q)$ with the same value
$\theta(\alpha)$. However, as for all $\theta(\alpha)$, their
limit is $\lim\limits_{\alpha\rightarrow 0}|\theta(\alpha)|^{2}=1$
or equivalently $\lim\limits_{\alpha\rightarrow
0}|\theta(\alpha)|=1$, in going to the low-energy limit each wave
function $\psi(q)$ is simply multiplied by the phase factor
$\theta(0)$. As a result we have Hilbert Space wave functions in
QM. Comparison of both pictures (Neumann's and Shr{\"o}dinger's)
is indicative of the fact that unitarity means the retention of
the probabilities $\omega_{i}(\alpha)$ or retention of the squared
modulus (and hence the modulus) for the function $\theta(\alpha)$:
$|\theta(\alpha)|^{2}$,($|\theta(\alpha)|$).That is
\\
\\
$$\frac{d\omega_{i}[\alpha(t)]}{dt}=0$$ or
$$\frac{d|\theta[\alpha(t)]|}{dt}=0.$$
\\
\\
In this way a set of unitary transformations of QMFL includes a
group $U$ of the unitary transformations for the wave functions
$\psi(q)$ in QM.
\\It is seen that on going from Planck's scale to the
conventional one , i.e. on transition from the Early Universe to
the current one, the scale has been rapidly changing in the
process of inflation expansion and the above conditions failed to
be fulfilled:
\begin{equation}\label{U26h}
\frac{d\omega_{i}[\alpha(t)]}{dt}\neq 0, {\sloppy}
\frac{d|\theta[\alpha(t)]|}{dt}\neq 0.
\end{equation}
In terms of the density pro-matrices of sections 2,3 this is a
limiting transition from the density pro-matrix in QMFL
$\rho(\alpha)$,$\alpha>0$ , that is a prototype of the pure state
at $\alpha\rightarrow 0$ to the density matrix $\rho(0)=\rho$
representing a pure state in QM. Mathematically this means that a
nontotal probability (below 1) is changed by the total one (equal
to 1). For the wave functions in Schr{\"o}dinger picture this
limiting transition from QMFL to QM is as follows:
\\
\\
$$\lim\limits_{\alpha\rightarrow 0}\theta(\alpha)\psi(q)=\psi(q)$$
up to the phase factor.
\\It is apparent that the above transition from QMFL to QM is
not a unitary process, as indicated in
\cite{shalyt1}-\cite{shalyt5} and section 3.2. However, the
unitarity may be recovered when we consider in a sense a reverse
process: absorption of the matter by a black hole and its
transition to singularity conforming to the reverse and nonunitary
transition from QM to QMFL. Thus, nonunitary transitions occur in
this picture twice:
\\
\\
$$I.(QMFL,OS,\alpha\approx 1/4)\stackrel{Big\enskip
Bang}{\longrightarrow}(QM,\alpha\approx 0)$$
\\
\\
$$II.(QM,\alpha\approx 0)\stackrel{absorbing\enskip BH
}{\longrightarrow}(QMFL,SBH,\alpha\approx 1/4).$$
\\
\\
Here the following abbreviations are used: OS for the Origin
Singularity; BH for a Black Hole; SBH for the Singularity in Black
Hole.
\\
As a result of these two nonunitary transitions, the total
unitarity may be recovered:
\\
\\
$$III.(QMFL,OS,\alpha\approx
1/4){\longrightarrow}(QMFL,SBH,\alpha\approx 1/4).$$
\\
\\
In such a manner the total information quantity in the Universe
remains unchanged, i.e. no information loss occurs.
\\ In terms of the deformed Liouville equation \cite{shalyt3}-\cite{shalyt5}
and section 4.1 we arrive to the expression with the same
right-hand parts for $t_{initial}\sim t_{Planck}$ and $t_{final}$
(for $\alpha\approx 1/4$).
\begin{eqnarray}\label{U27h}
\frac{d\rho[\alpha(t),t]}{dt}=\sum_{i}
\frac{d\omega_{i}[\alpha(t)]}{dt}|i(t)><i(t)|-\nonumber \\
-i[H,\rho(\alpha)]= d[ln\omega(\alpha)]\rho
(\alpha)-i[H,\rho(\alpha)].
\end{eqnarray}
It should be noted that for the closed Universe one can consider
Final Singularity (FS) rather than the Singularity of Black Hole
(SBH), and then the right-hand parts of diagrams II and III will
be changed:
\\
\\
$$IIa.(QM,\alpha\approx 0)\stackrel{Big\enskip Crunch
}{\longrightarrow}(QMFL,FS,\alpha\approx 1/4),$$
\\
\\
$$IIIa.(QMFL,OS,\alpha\approx
1/4){\longrightarrow}(QMFL,FS,\alpha\approx 1/4)$$
\\
\\
At the same time, in this case the general unitarity and
information are still retained, i.e. we again have the "unitary"
product of two "nonunitary" arrows:
\\
\\
$$IV.(QMFL,OS,\alpha\approx 1/4)\stackrel{Big\enskip
Bang}{\longrightarrow}(QM,\alpha\approx 0)\stackrel{Big\enskip
Crunch }{\longrightarrow}(QMFL,FS,\alpha\approx 1/4).$$
\\
\\
Finally, arrow III may appear directly, i.e. without the
appearance of arrows I è II, when in the Early Universe mini BH
are arising:
\\
\\
$$IIIb.(QMFL,OS,\alpha\approx 1/4){\longrightarrow}(QMFL,
mini\enskip BH, SBH,\alpha\approx 1/4).$$
\\
\\
Note that here, unlike the previous cases, a unitary transition
occurs immediately, without any additional nonunitary ones, and
with retention of the total information.
\\ Another approach to the information paradox problem associated
with the above-mentioned methods (density matrix deformation) is
the introduction and investigation of a new value, namely entropy
density per minimum unit area. This approach is described in the
following subsection.

\subsection{ Entropy density matrix and information loss problem }

In \cite{shalyt1}-\cite{shalyt5} the authors were too careful,
when introducing for density pro-matrix $\rho(\alpha)$ the value
$S_{\alpha}$ generalizing the ordinary statistical entropy:
\\
 $$S_{\alpha}=-Sp[\rho(\alpha)\ln(\rho(\alpha))]=
 -<\ln(\rho(\alpha))>_{\alpha}.$$
\\
In \cite{shalyt4},\cite{shalyt5} it was noted that $S_{\alpha}$
means the entropy density   on a minimum unit area depending on
the scale. In fact a more general concept accepts the form of the
entropy density matrix \cite{shalyt11}:
\begin{equation}\label{U4h}
S^{\alpha_{1}}_{\alpha_{2}}=-Sp[\rho(\alpha_{1})\ln(\rho(\alpha_{2}))]=
-<\ln(\rho(\alpha_{2}))>_{\alpha_{1}},
\end{equation}
where $0< \alpha_{1},\alpha_{2}\leq 1/4.$
\\ $S^{\alpha_{1}}_{\alpha_{2}}$ has a clear physical meaning:
the entropy density is computed  on the scale associated with the
deformation parameter $\alpha_{2}$ by the observer who is at a
scale corresponding to the deformation parameter $\alpha_{1}$.
Note that with this approach the diagonal element
$S_{\alpha}=S_{\alpha}^{\alpha}$,of the described matrix
$S^{\alpha_{1}}_{\alpha_{2}}$ is the density of entropy, measured
by the observer  who is at the same scale  as the measured object
associated with the deformation parameter $\alpha$. In \cite{r15}
and section 4.3 such a construction was used implicitly in
derivation of the semiclassical Bekenstein-Hawking formula for the
Black Hole entropy:

a) for the initial (approximately pure) state
\\
$$S_{in}=S_{0}^{0}=0,$$
\\
b) using the exponential ansatz(\ref{U15}),we obtain:
\\
$$S_{out}=S^{0}_{\frac{1}{4}}=-<ln[exp(-1/4)]\rho_{pure}>=-<\ln(\rho(1/4))>
=\frac{1}{4}.$$
\\
So increase in the entropy density for an external observer at the
large-scale limit is 1/4. Note that increase of the entropy
density(information loss) for the observer crossing the horizon of
the black hole's events and moving with the information flow to
singularity will be smaller:
\\
$$S_{out}=S_{\frac{1}{4}}^{\frac{1}{4}}=-Sp(exp(-1/4)
ln[exp(-1/4)]\rho_{pure})=-<\ln(\rho(1/4))>_{\frac{1}{4}} \approx
0.1947 .$$ It is clear that this fact may be interpreted as
follows: for the observer moving together with information its
loss can  occur only at the transition to smaller scales, i.e. to
greater deformation parameter $\alpha$. \\
\\ Now we consider the general Information Problem.
Note that with the well-known Quantum Mechanics (QM) the entropy
density matrix $S^{\alpha_{1}}_{\alpha_{2}}$ (\ref{U4h}) is
reduced only to one element $S_{0}^{0}$ . So we can not test
anything. Moreover, in previous works relating the quantum
mechanics of black holes and information paradox
\cite{r15},\cite{r13},\cite{r16} the initial and final states when
a particle hits the hole are treated proceeding from different
theories (QM and QMFL respectively), as was indicated in diagram
II:
\\
\\
(Large-scale limit, QM,
 density matrix) $\rightarrow$ (Black Hole, singularity, QMFL,
density pro-matrix).
\\
\\
Of course in this case any conservation of information is
impossible as these theories are based on different concepts of
entropy. Simply saying, it is incorrect to compare the entropy
interpretations of two different theories (QM and QMFL)where this
notion is originally differently understood. So the chain above
must be symmetrized by accompaniment of the arrow on the left ,so
in an ordinary situation we have a chain (diagram III):
\\
\\
(Early Universe, origin singularity, QMFL, density pro-matrix)
$\rightarrow$
\\ (Large-scale limit, QM,
 density matrix)$\rightarrow$ (Black Hole, singularity, QMFL,
density pro-matrix).
\\
\\
So it's more correct to compare entropy close to the origin and
final (Black hole) singularities. In other words, it is necessary
to take into account not only the state, where information
disappears, but also that whence it appears. The question arises,
whether in this case the information is lost for every separate
observer. For the event under consideration this question sounds
as follows: are the entropy densities S(in) and S(out) equal for
every separate observer? It will be shown that in all conceivable
cases they are equal.

1) For the observer in the large-scale limit (producing
measurements in the semiclassical approximation) $\alpha_{1}=0$
\\
\\
$S(in)=S^{0}_{\frac{1}{4}}$ (Origin singularity)
\\
\\
$S(out)=S^{0}_{\frac{1}{4}}$ (Singularity in Black Hole)
\\
\\
So $S(in)=S(out)=S^{0}_{\frac{1}{4}}$. Consequently, the initial
and final densities of entropy are equal and there is no
information loss.
\\
2) For the observer moving together with the information flow in
the general situation  we have the chain:
\\
$$S(in)\rightarrow S(large-scale)\rightarrow S(out),$$
\\
where $S(large-scale)=S^{0}_{0}=S$. Here $S$ is an ordinary
entropy of Quantum Mechanics(QM), but
$S(in)=S(out)=S^{\frac{1}{4}}_{\frac{1}{4}}$,- value considered in
QMFL. So in this case the initial and final densities of entropy
are equal without any loss of information.
\\
3) This is a special case of 2), when we do not leave out of the
Early Universe considering the processes with the participation of
black mini-holes only. In this case the originally specified chain
becomes shorter by one section (diagram IIIb):
\\
\\
(Early Universe, origin singularity, QMFL, density
pro-matrix)$\rightarrow$ (Black Mini-Hole, singularity, QMFL,
density pro-matrix),
\\
\\
and member $S(large-scale)=S^{0}_{0}=S$ disappears at the
corresponding chain of the entropy density associated with the
large-scale:
\\
$$S(in)\rightarrow S(out),$$
\\
It is, however, obvious that in case
$S(in)=S(out)=S^{\frac{1}{4}}_{\frac{1}{4}}$ the density of
entropy is preserved. Actually this event was mentioned in
\cite{shalyt5},where from the basic principles it has been found
that black mini-holes do not radiate, just in agreement with the
results of other authors \cite{r17},\cite{r31},\cite{r30}.
\\ As a result, it's possible to write briefly
\\
$$S(in)=S(out)=S^{\alpha}_{\frac{1}{4}},$$
\\
where $\alpha$ - any value in the interval $0<\alpha\leq 1/4.$
\\ Actually our inferences are similar to those of  section 4.1
in terms of the Liouville's equation deformation:
\\
$$\frac{d\rho}{dt}=\sum_{i}
\frac{d\omega_{i}[\alpha(t)]}{dt}|i(t)><i(t)|-i[H,\rho(\alpha)]=
\\d[ln\omega(\alpha)]\rho (\alpha)-i[H,\rho(\alpha)].$$
\\
The main result of this section is a necessity to account for the
member $d[ln\omega(\alpha)]\rho (\alpha)$,deforming the right-side
expression of $\alpha\approx 1/4$.

\subsection {Unitarity, non-unitarity and Heisenberg's algebra
deformation}

The above-mentioned unitary and nonunitary transitions may be
described in terms of Heisenberg's algebra deformation
(deformation of commutators) as well. We use the principal results
and designations from \cite{Magg}.In the process the following
assumptions are resultant: 1)The three-dimensional rotation group
is not deformed; angular momentum ${\bf J}$ satisfies the
undeformed $SU(2)$ commutation relations, whereas the coordinate
and momenta satisfy the undeformed commutation relations $\left[
J_i,x_j\right] =i\epsilon_{ijk}x_k, \left[ J_i,p_j\right]
=i\epsilon_{ijk}p_k$. 2) The momenta commute between themselves:
$\left[ p_i,p_j\right] =0$, so the translation group is also not
deformed. 3) Commutators $\left[ x,x\right]$ and $\left[
x,p\right]$ depend on the deformation parameter $\kappa$ with the
dimension of mass. In the limit $\kappa\rightarrow \infty$ with
$\kappa$ much larger than any energy the canonical commutation
relations are recovered.
\\
For a specific realization of points 1) to 3) the generating GUR
are of the form \cite{Magg}: ($\kappa$-deformed Heisenberg
algebra)
\begin{eqnarray}
\left[ x_i ,x_j \right] &= & -\frac{\hbar^2}{\kappa^2}\,
i\epsilon_{ijk}J_k\label{xx}\\ \left[ x_i , p_j \right]   &= &
i\hbar\delta_{ij} (1+\frac{E^2}{\kappa^2})^{1/2}\, .\label{xp}
\end{eqnarray}
Here $E^2=p^2+m^2$. Note that in this formalism the transition
from GUR to UR, or equally from QMFL to QM with $\kappa\rightarrow
\infty$ or from Planck scale to the conventional one, is
nonunitary exactly following the transition from density
pro-matrix to the density matrix in previous sections:
\\
$$\rho(\alpha\neq 0)\stackrel{\alpha\rightarrow
0}{\longrightarrow}\rho.$$
\\
Then the first arrow I in the formalism of this section may be as
follows:
\\
$$I^{\prime}.(GUR,OS,\kappa\sim M_{p})\stackrel{Big\enskip
Bang}{\longrightarrow}(UR,\kappa=\infty)$$ or what is the same
$$I^{\prime\prime}.(QMFL,OS,\kappa\sim M_{p})\stackrel{Big\enskip
Bang}{\longrightarrow}(QM,\kappa=\infty),$$
\\
where $M_{p}$ is the Planck mass.
\\
\\In some works of the last two
years Quantum Mechanics for a Black Hole has been already
considered as a Quantum Mechanics with GUR \cite{r17},\cite{r31}.
As a consequence, by this approach the Black Hole is not
completely evaporated but rather some stable remnants always
remain in the process of its evaporation with a mass $\sim M_{p}$.
In terms of \cite{Magg} this means nothing else but a reverse
transition: $(\kappa=\infty)\rightarrow(\kappa\sim M_{p})$. And
for an outside observer this transition is of the form:
\\
$$II^{\prime}.(UR,\kappa=\infty)\stackrel{absorbing\enskip
BH}{\longrightarrow}(GUR,SBH,\kappa\sim M_{p}),$$ that is
$$II^{\prime\prime}.(QM,\kappa=\infty)\stackrel{absorbing\enskip
BH}{\longrightarrow}(QMFL,SBH,\kappa\sim M_{p}).$$
\\
\\
So similar to the previous section, two nonunitary inverse
transitions a)$I^{\prime},(I^{\prime\prime})$ and
b)$II^{\prime},(II^{\prime\prime})$ are liable to generate a
unitary transition:
\\
$$III^{\prime}.(GUR,OS,\kappa\sim M_{p})\stackrel{Big\enskip
Bang}{\longrightarrow}(UR,\kappa=\infty)\stackrel{absorbing\enskip
BH}{\longrightarrow}(GUR,SBH,\kappa\sim M_{p}),$$
\\
or to summerize
\\
$$III^{\prime\prime}.(GUR,OS,\kappa\sim
M_{p})\rightarrow(GUR,SBH,\kappa\sim M_{p})$$
\\
In conclusion of this section it should be noted that an approach
to the Quantum Mechanics at Planck Scale using the Heisenberg
algebra deformation (similar to the approach based on the density
matrix deformation from the  section3) gives a deeper insight into
the possibility of retaining the unitarity and the total quantity
of information in the Universe, making possible the solution of
Hawking's Information Paradox Problem
\cite{r15},\cite{r13},\cite{r16}.

\subsection {Statistical mechanics deformation and transitions}
Naturally, deformation of Quantum Mechanics in the Early Universe
is associated with the Statistical Mechanics deformation as
indicated in \cite{shalyt6,shalyt7}. In case under consideration
this simply implies a transition from the Generalized Uncertainty
Relations (GUR) of Quantum Mechanics to GUR in Thermodynamics
\cite{shalyt7},\cite{shalyt9,shalyt10}. The latter are
distinguished from the normal uncertainty relations by:
\begin{equation}\label{U1T}
\Delta \frac{1}{T}\geq\frac{k}{\Delta U},
\end{equation}
i.e. by inclusion of the high-temperature term into the right-hand
side (section 6 of this chapter)
\begin{equation}\label{U2T}
\Delta \frac{1}{T}\geq
  \frac{k}{\Delta U}+\alpha^{\prime}
  \frac{1}{T_{p}^2}\frac{\Delta U}{k}+...     .
\end{equation}
Thus, denoting the Generalized Uncertainty Relations in
Thermodynamics as GURT and using abbreviation URT for the
conventional ones, we obtain a new form of diagram I from section
III ($I^{\prime}$ of section IV respectively):
\\
 $$I^{T}.(GURT,OS)\stackrel{Big\enskip
Bang}{\longrightarrow}(URT).$$
\\
In \cite{shalyt6,shalyt7} and section 5 of this chapter the
Statistical Mechanics deformation associated with GURT is
implicitly assumed by the introduction of the respective
deformation for the statistical density matrix $\rho_{stat}(\tau)$
where $0<\tau \leq 1/4$. Obviously, close to the Origin
Singularity $\tau\approx 1/4$. Because of this, arrow $I^{T}$ may
be represented in a more general form as
\\
 $$I^{Stat}.(GURT,OS,\rho_{stat}(\tau),\tau\approx 1/4)
 \stackrel{Big\enskip
Bang}{\longrightarrow}(URT,\rho_{stat},\tau\approx 0).$$
\\
The reverse transition is also possible. In \cite{r17},\cite{r31}
it has bee shown that Statistical Mechanics of Black Hole should
be consistent with the deformation of a well-known Statistical
Mechanics. The demonstration of an *upper* bound for temperature
in Nature, given by Planck temperature and related to Black Hole
evaporation, was provided in \cite{Castro3}. It is clear that
emergence of such a high temperatures is due to GURT. And we have
the following diagram that is an analog of diagrams II and
$II^{\prime}$ for Statistical Mechanics:
\\
$$II^{Stat}.(URT,\rho_{stat},\tau\approx 0)
\stackrel{absorbing\enskip
BH}{\longrightarrow}(GURT,SBH,\rho_{stat}(\tau),\tau\approx 1/4).
$$
\\
By this means, combining $I^{Stat}$ and $II^{Stat}$, we obtain
$III^{Stat}$ representing a complete statistical-mechanics analog
for quantum-mechanics diagrams $III$ and $III^{\prime}$:
\\
$$III^{Stat}.(GURT,OS,\tau\approx 1/4)
 \stackrel{Big\enskip
Bang,\enskip absorbing\enskip BH}{\longrightarrow} (GURT,
SBH,\tau\approx 1/4).$$
\\
And in this case two nonunitary transitions $I^{Stat}$ and
$II^{Stat}$ in the end lead to a unitary transition $III^{Stat}$.

\section{The Universe as a Nonuniform Lattice in Finite-Volume
Hypercube}
 In this section a new small parameter associated with
the density matrix deformation (density pro-matrix)studied in
previous sections  is introduced into the Generalized Quantum
Mechanics (GQM), i.e. quantum mechanics involving description of
the Early Universe. It is noted that this parameter has its
counterpart in the Generalized Statistical Mechanics. Both
parameters offer a number of merits: they are dimensionless,
varying over the interval from 0 to 1/4 and assuming in this
interval a discrete series of values. Besides, their definitions
contain all the fundamental constants. These parameters are very
small for the conventional scales and temperatures, e.g. the value
of the first parameter is on the order of $\approx 10^{-66+2n}$,
where $10^{-n}$ is the measuring scale and the Planck scale $\sim
10^{-33}cm$ is assumed. The second one is also too small for the
conventional temperatures, that is those much below the Planck's.
It is demonstrated that relative to the first of these parameters
the Universe may be considered as a nonuniform lattice in the
four-dimensional hypercube with dimensionless finite-length (1/4)
edges. And the time variable is also described by one of the
above-mentioned dimensions due to the second parameter and
Generalized Uncertainty Relation in thermodynamics. In this
context the lattice is understood as a deformation rather than
approximation \cite{shalyt14}.

\subsection{Definition of lattice}

It should be noted that according to subsection 3.2 a minimum
measurable length is equal
 to $l^{*}_{min}=2l_{min}$ being a nonreal
number at point $l_{min}$,$Sp[\rho(\alpha)]$. Because of this, a
space part of the Universe is a lattice with a spacing of
$a_{min}=2l_{min}\sim 2l_{p}$. In consequence the first issue
concerns the lattice spacing of any lattice-type model(for example
\cite{rLat1,rLat2}): a selected lattice spacing $a_{lat}$ should
not be less than $a_{min}$,i.e. always $a_{lat}\geq a_{min}>0$.
Besides, a continuum limit in any lattice-type model is meaning
$a_{lat}\rightarrow a_{min}>0$ rather than $a_{lat}\rightarrow 0$.
\\ Proceeding from $\alpha$, for each space dimension we have a
discrete series of rational values for the inverse squares of even
numbers nonuniformly distributed along the real number line
$\alpha = 1/4, 1/16,1/36,1/64,...$. A question arises,is this
series somewhere terminated or, on the contrary, is it infinite?
The answer depends on the answers to two other questions:
\\(1) Is there theoretically a maximum measurability limit for the scales
$l_{max}$?  and
\\(2) Is our Universe closed in the sense that its
extension may be sometime replaced by compression, when a maximum
extension precisely gives a maximum scale $l_{max}$?
\\Should an answer to one of these questions be positive, we should have
$0<l^{2}_{min}/l^{2}_{max}\leq\alpha\leq1/4$ rather than condition
1 of {\bf Definition 1, subsection 3.1 of this chapter}
\\ Note that in the majority of cases all three space dimensions are
equal, at least at large scales, and hence their associated values
of $\alpha$ parameter should be identical. This means that for
most cases, at any rate in the large-scale (low-energy) limit, a
single deformation parameter $\alpha$ is sufficient to accept one
and the same value for all three dimensions to a high degree of
accuracy. In the general case, however, this is not true, at least
for very high energies (on the order of the Planck's), i.e. at
Planck scales, due to noncommutativity of the spatial coordinates
\cite{r4},\cite{Magg}:
\\
$$\left[ x_i ,x_j \right]\neq 0.$$
\\
In consequence in the general case we have a point with
coordinates ${\bf
\widetilde{\alpha}}=(\alpha_{1},\alpha_{2},\alpha_{3})$ in the
normal(three-dimensional) cube $I_{1/4}^{3}$ of side
$I_{1/4}=(0;1/4]$.
\\ It should be noted that this universal cube may be extended to
the four-dimensional hypercube by inclusion of the additional
parameter $\tau,\tau\in I_{1/4}$ that is generated by internal
energy of the statistical ensemble and its temperature for the
events when this notion is the case. It will be recalled that
$\tau$ parameter occurs from a maximum temperature that is  in its
turn generated by the Generalized Uncertainty Relations of "energy
– time" pair in GUR (see {\bf Definition 2} in subsection 5.1 and
\cite{shalyt6,shalyt7}).
\\So $\tau$ is a counterpart (twin) of $\alpha$, yet for the Statistical
Mechanics. At the same time, originally for $\tau$ nothing implies
the discrete properties of parameter $\alpha$ indicated above:
\\ for $\tau$ there is a discrete series (lattice) of the rational
values of inverse squares for even numbers not uniformly
distributed along the real number line: $\tau = 1/4, 1/16,
1/36,1/64,...$.
 \\ Provided such a series exists actually,
\\The finitness and infinity question for this series amounts to
two other questions:
\\(1) Is there theoretically any minimum measurability limit for
the average temperature of the Universe $T_{min}\neq 0$ and
\\(2) Is our Universe closed in a sense that its extension may be sometime
replaced by compression? Then maximum extension just gives a
minimum temperature $T_{min}\neq 0$.
\\ The question concerning the discretization of parameter $\tau$
is far from being idle. The point is that originally by its nature
this parameter seems to be continuous as it is associated with
temperature. Nevertheless, in the following section we show that
actually $\tau$ is dual in nature: it is directly related to time
that is in turn quantized,in the end giving a series $\tau = 1/4,
1/16, 1/36,1/64,...$.

\subsection{Dual nature of parameter $\tau$ and its temporal
aspect}

In this way when at point ${\bf\widetilde{\alpha}}$ of the normal
(three-dimensional) cube $I_{1/4}^{3}$ of side $I_{1/4}=(0;1/4]$
an additional "temperature" variable $\tau$ is added, a nonuniform
lattice of the point results, where we denote
$\widetilde{\alpha}_{\tau}=(\widetilde{\alpha},\tau)=
(\alpha_{1},\alpha_{2},\alpha_{3},\tau)$ at the four-dimensional
hypercube $I_{1/4}^{4}$, every coordinate of which assumes one and
the same discrete series of values: 1/4, 1/16, 1/36,1/64,...,
$1/4n^{2}$,... .(Further it is demonstrated that $\tau$ is also
taking on a discrete series of values.) The question arises,
whether time "falls" within this picture. The answer is positive.
Indeed, parameter $\tau$ is dual (thermal and temporal) in nature
owing to introduction of the Generalized Uncertainty Relations in
Thermodynamics (GURT)
(\cite{shalyt9},\cite{shalyt10},\cite{shalyt7} and section 6):
\\
$$\Delta \frac{1}{T} \geq
  \frac{k}{\Delta U}+\alpha^{\prime}
  \frac{1}{T_{p}^2}\,
  \frac{\Delta U}{k}+...,$$
\\
where $k$ - Boltzmann constant, $T$ - temperature of the ensemble,
$U$ - its internal energy. A direct implication of the latter
inequality is occurrence of a "maximum" temperature $T_{max}$ that
is inversely proportional to "minimal" time $t_{min}\sim t_{p}$:
\\
$$T_{max}=\frac{\hbar}{2\surd \alpha^{\prime}t_{p}
k}=\frac{\hbar}{\Delta t_{min} k}$$
\\
However, $t_{min}$ follows from the Generalized Uncertainty
Relations in Quantum Mechanics for "energy-time" pair
(\cite{shalyt6},\cite{shalyt7} and section 5):
\\
$$\Delta t\geq\frac{\hbar}{\Delta
E}+\alpha^{\prime}t_{p}^2\,\frac{\Delta E}{ \hbar}.$$
\\
Thus, $T_{max}$ is the value relating GUR and GURT together (see
sections 5,6 and \cite{shalyt9},\cite{shalyt10},\cite{shalyt7})
\begin{equation}\label{U18L}
\left\{
\begin{array}{lll}
\Delta x & \geq & \frac{\displaystyle\hbar}{\displaystyle\Delta
p}+\alpha^{\prime} L_{p}^2\,\frac{\displaystyle\Delta
p}{\displaystyle\hbar}+... \\
  &  &  \\
  \Delta t & \geq &  \frac{\displaystyle\hbar}{\displaystyle\Delta E}+\alpha^{\prime}
  t_{p}^2\,\frac{\displaystyle\Delta E}{\displaystyle\hbar}+... \\
  &  &  \\

  \Delta \frac{\displaystyle 1}{\displaystyle T} & \geq &
  \frac{\displaystyle k}{\displaystyle\Delta U}+\alpha^{\prime}
  \frac{\displaystyle 1}{\displaystyle T_{p}^2}\,
  \frac{\displaystyle\Delta U}{\displaystyle k}+...,
\end{array} \right.
\end{equation},
since the thermodynamic value $T_{max}$ (GURT) is associated with
the quantum-mechanical one $E_{max}$ (GUR) by the formula from
section 5:
\\
$$T_{max}=\frac{E_{max}}{k}$$
\\
The notion of value $t_{min}\sim 1/T_{max}$ is physically crystal
clear, it means a minimum time for which any variations in the
energy spectrum of every physical system may be recorded.
Actually, this value is equal to $t^{*}_{min}=2t_{min}\sim t_{p}$
as at the initial points $l_{min}$ and $T_{max}$ the spurs of the
quantum-mechanical and statistical density pro-matrices
${\bf\rho_(\alpha)}$ and ${\bf\rho_{stat}(\tau)}$ are complex,
determined only beginning from $2l_{min}$ è
$T^{*}_{max}=\frac{1}{2}T_{max}$ \cite{shalyt5},\cite{shalyt7}
that is associated with the same time point
$t^{*}_{min}=2t_{min}$. For QMFL this has been noted in the
previous section.
\\  In such a manner a discrete series $l^{*}_{min},2l^{*}_{min}$,...
generates in QMFL the discrete time series
$t^{*}_{min},2t^{*}_{min},...$, that is in turn associated (due to
GURT)with a discrete temperature series
$T^{*}_{max}$,$\frac{1}{2}T^{*}_{max}$, ... . From this it is
inferred that a "temperature" scale $\tau$ may be interpreted as a
"temporal" one $\tau=t_{min}^{2}/t^{2}$. In both cases the
generated series has one and the same discrete set of values for
parameter $\tau$ :$\tau = 1/4, 1/16, 1/36,1/64,..., 1/4n^{2}$,...
. Thus, owing to time quantization in QMFL, one is enabled to
realize quantization of temperature in the generalized Statistical
Mechanics with the use of GURT.
\\Using $Lat_{\widetilde{\alpha}}$, we denote the lattice in cube
$I_{1/4}^{3}$ formed by points $\widetilde{\alpha}$, and through
$Lat^{\tau}_{\widetilde{\alpha}}$ we denote the lattice in
hypercube $I_{1/4}^{4}$ that is formed by points
$\widetilde{\alpha}_{\tau}=(\widetilde{\alpha},\tau)$.
\subsection{Quantum theory \\ for the lattice in hypercube}

Any quantum theory may be defined for the indicated lattice in
hypercube.
 To this end, we recall the principal result of subsection 4.4 as
{\bf Definition $1^{\prime}$} in this section with $\alpha$
changed by $\widetilde{\alpha}$:
\\ \noindent {\bf Definition
$1^{\prime}$} {\bf Quantum Mechanics with Fundamental Length}
\\ {\bf (Shr{\"o}dinger's picture)}
\\
Here, the prototype of Quantum Mechanical normed wave function (or
the pure state prototype) $\psi(q)$ with $\int|\psi(q)|^{2}dq=1$
in QMFL is
$\psi(\widetilde{\alpha},q)=\theta(\widetilde{\alpha})\psi(q)$.
The parameter of deformation $\widetilde{\alpha}\in I_{1/4}^{3}$.
Its properties are
$|\theta(\widetilde{\alpha})|^{2}<1$,$\lim\limits_{|\widetilde{\alpha}|\rightarrow
0}|\theta(\widetilde{\alpha})|^{2}=1$ and the relation
$|\theta(\alpha_{i})|^{2}-|\theta(\alpha_{i})|^{4}\approx
\alpha_{i}$ takes place. In such a way the total probability
always is less than 1:
$p(\widetilde{\alpha})=|\theta(\widetilde{\alpha})|^{2}
=\int|\theta(\widetilde{\alpha})|^{2}|\psi(q)|^{2}dq<1$ tending to
1, when  $\|\widetilde{\alpha}\|\rightarrow 0$. In the most
general case of the arbitrarily normed state in QMFL(mixed state
prototype)
$\psi=\psi(\widetilde{\alpha},q)=\sum_{n}a_{n}\theta_{n}(\widetilde{\alpha})\psi_{n}(q)$
with $\sum_{n}|a_{n}|^{2}=1$ the total probability is
$p(\widetilde{\alpha})=\sum_{n}|a_{n}|^{2}|\theta_{n}(\widetilde{\alpha})|^{2}<1$
and
 $\lim\limits_{\|\widetilde{\alpha}\|\rightarrow 0}p(\widetilde{\alpha})=1$.

It is natural that Shr{\"o}dinger equation is also deformed in
QMFL. It is replaced by the equation

\begin{equation}\label{U24L}
\frac{\partial\psi(\widetilde{\alpha},q)}{\partial t}
=\frac{\partial[\theta(\widetilde{\alpha})\psi(q)]}{\partial
t}=\frac{\partial\theta(\widetilde{\alpha})}{\partial
t}\psi(q)+\theta(\widetilde{\alpha})\frac{\partial\psi(q)}{\partial
t},
\end{equation}
where the second term in the right-hand side generates the
Shr{\"o}dinger equation as
\begin{equation}\label{U25L}
\theta(\widetilde{\alpha})\frac{\partial\psi(q)}{\partial
t}=\frac{-i\theta(\widetilde{\alpha})}{\hbar}H\psi(q).
\end{equation}
Here $H$ is the Hamiltonian and the first member is added
similarly to the member that appears in the deformed Liouville
equation, vanishing when $\theta[\widetilde{\alpha}(t)]\approx
const$. In particular, this takes place in the low energy limit in
QM, when $\|\widetilde{\alpha}\|\rightarrow 0$. It should be noted
that the above theory is not a time reversal of QM because the
combination $\theta(\widetilde{\alpha})\psi(q)$ breaks down this
property in the deformed Shr{\"o}dinger equation. Time-reversal is
conserved only in the low energy limit, when a quantum mechanical
Shr{\"o}dinger equation is valid.
\\ According to {\bf Definition $1^{\prime}$}everywhere $q$ is
the coordinate of a point at the three-dimensional space. As
indicated in \cite{shalyt1}--\cite{shalyt5} and section 3.2, for a
density pro-matrix there exists an exponential ansatz satisfying
the formula (\ref{U13}) of {\bf Definition 1}, section 3.1:
\begin{equation}\label{U26L}
\rho^{*}(\alpha)=\sum_{i}\omega_{i} exp(-\alpha)|i><i|,
\end{equation}
where all $\omega_{i}>0$ are independent of $\alpha$  and their
sum is equal to 1. In this way
$Sp[\rho^{*}(\alpha)]=exp(-\alpha)$. Then in the momentum
representation $\alpha=p^{2}/p^{2}_{max}$, $p_{max}\sim
p_{pl}$,where $p_{pl}$ is the Planck momentum. When present in
matrix elements, $exp(-\alpha)$  damps the contribution of great
momenta in a perturbation theory.
\\ It is clear that for each of the coordinates $q_{i}$ the
exponential ansatz makes a contribution to the deformed wave
function $\psi(\widetilde{\alpha},q)$ the modulus of which equals
$exp(-\alpha_{i}/2)$  and, obviously, the same contribution to the
conjugate function $\psi^{*}(\widetilde{\alpha},q)$. Because of
this, for exponential ansatz one may write
\begin{equation}\label{U27L}
\psi(\widetilde{\alpha},q)=\theta(\widetilde{\alpha})\psi(q),
\end{equation}
where $|\theta(\widetilde{\alpha})|=exp(-\sum_{i}\alpha_{i}/2)$.
As noted above, the last exponent of the momentum representation
reads $exp(-\sum_{i}p_{i}^{2}/2p_{max}^{2})$ and in this way it
removes UV (ultra-violet) divergences in the theory.

It follows that $\widetilde{\alpha}$ is a new small parameter.
Among its obvious advantages one could name:
\\1)  its dimensionless nature,
\\2)  its variability over the finite interval $0<\alpha_{i}\leq 1/4$.
Besides, for the well-known physics it is actually very small:
$\alpha\sim 10^{-66+2n}$, where $10^{-n}$ is the measuring scale.
Here the Planck scale $\sim 10^{-33}cm$ is assumed;
\\3)and finally the calculation of this parameter involves all
three fundamental constants, since by Definition 1 of subsection
3.1 $\alpha_{i}= l_{min}^{2}/x_{i}^{2 }$, where $x_{i}$ is the
measuring scale on i-coordinate and $l_{min}^{2}\sim
l_{pl}^{2}=G\hbar/c^{3}$.
\\ Therefore, series expansion in $\alpha_{i}$ may be of great importance.
Since all the field components and hence the Lagrangian will be
dependent on $\widetilde{\alpha}$, i.e.
$\psi=\psi(\widetilde{\alpha}),L=L(\widetilde{\alpha})$, quantum
theory may be considered as a theory of lattice
$Lat_{\widetilde{\alpha}}$ and hence of lattice
$Lat^{\tau}_{\widetilde{\alpha}}$.
\subsection{Introduction of
quantum field theory and initial analysis} With the use of this
approach for the customary energies a Quantum Field Theory (QFT)
is introduced with a high degree of accuracy. In our context
"customary" means the energies much lower than the Planck ones.
\\ It is important that as the spacing of lattice
 $Lat^{\tau}_{\widetilde{\alpha}}$
is decreasing in inverse proportion to the square of the
respective node, for a fairly large node number $N>N_{0}$  the
lattice edge beginning at this node $\ell_{N,N+1}$
\cite{shalyt1}--\cite{shalyt5} will be of length $\ell_{N,N+1}\sim
1/4N^{3}$, and by this means edge lengths of the lattice are
rapidly decreasing with the spacing number. Note that in the
large-scale limit this (within any preset accuracy)leads to
parameter $\alpha=0$, pure states and in the end to QFT. In this
way a theory for the above-described lattice presents a
deformation of the originally continuous variant of this theory as
within the developed approach continuity is accurate to $\approx
10^{-66+2n}$, where $10^{-n}$ is the measuring scale and the
Planck scale $\sim 10^{-33}cm$ is assumed. Whereas the lattice per
se $Lat^{\tau}_{\widetilde{\alpha}}$ may be interpreted as a
deformation of the space continuum with the deformation parameter
equal to the varying edge length
$\ell_{\alpha^{1}_{\tau_{1}},\alpha^{2}_{\tau_{2}}}$, where
$\alpha^{1}_{\tau_{1}}$ è $\alpha^{2}_{\tau_{2}}$ are two adjacent
points of the lattice $Lat^{\tau}_{\widetilde{\alpha}}$.
Proceeding from this, all well-known theories including
$\varphi^{4}$, QED, QCD and so on may be studied based on the
above-described lattice.
\\ Here it is expedient to make the following remarks:
\\{\bf (1) going on from the well-known energies of these theories
to higher energies (UV behavior) means a change from description
of the theory's behavior for the lattice portion with high edge
numbers to the portion with low numbers of the edges;
\\ (2) finding of quantum correction factors for the primary deformation
parameter $\widetilde{\alpha}$ is a power series expansion in each
$\alpha_{i}$. In particular, in the simplest case (Definition
$1^{\prime}$ of subsection 8.3 )means expansion of the left side
in relation
$|\theta(\alpha_{i})|^{2}-|\theta(\alpha_{i})|^{4}\approx
\alpha_{i}$:
\\
$$|\theta(\alpha_{i})|^{2}-|\theta(\alpha_{i})|^{4}=
\alpha_{i}+a_{0}\alpha^{2}_{i}+a_{1}\alpha^{3}_{i}+...$$
\\
and calculation of the associated coefficients $a_{0},a_{1},...$.}
This approach to calculation of the quantum correction factors may
be used in the formalism for density pro-matrix (Definition 1 of
subsection 3.1). In this case, the primary relation (\ref{U13}) of
{\bf Definition 1}, section 3.1 may be written in the form of a
series
\begin{equation}\label{U28L}
Sp[\rho(\alpha)]-Sp^{2}[\rho(\alpha)]=\alpha+a_{0}\alpha^{2}
+a_{1}\alpha^{3}+...     .
\end{equation}
As a result, a measurement procedure using the exponential ansatz
 may be understood as the calculation of factors
$a_{0}$,$a_{1}$,... or the definition of additional members in the
exponent "destroying" $a_{0}$,$a_{1}$,... \cite{shalyt13}. It is
easy to check that the exponential ansatz gives $a_{0}=-3/2$,
being coincident with the logarithmic correction factor for the
Black Hole entropy \cite{r22}.
\\ Most often a quantum theory is considered at zero temperature
$T=0$, in this context amounting to nesting of the
three-dimensional lattice $Lat_{\widetilde{\alpha}}$ into the
four-dimensional one:
$Lat^{\tau}_{\widetilde{\alpha}}$:$Lat_{\widetilde{\alpha}}\subset
Lat^{\tau}_{\widetilde{\alpha}}$ and nesting of the cube
$I_{1/4}^{3}$ into the hypercube $I_{1/4}^{4}$ as a bound given by
equation $\tau=0$. However, in the most general case the points
with nonzero values of $\tau$ may be important as there is a
possibility for nonzero temperature $T\neq0$ (quantum field theory
at finite temperature) that is related to the value of $\tau$
parameter, though very small but still nonzero: $\tau\neq0$. To
illustrate: in QCD for the normal lattice \cite{Di} a critical
temperature $T_{c}$ exists so that the following is fulfilled:
\\
at $$T<T_{c}$$ the confinement phase occurs,
\\
and for $$T>T_{c}$$ the deconfinement is the case.
\\ A critical temperature $T_{c}$ is associated with the "critical"
parameter $\tau_{c}=T^{2}_{c}/T^{2}_{max}$  and the selected bound
of hypercube $I_{1/4}^{4}$ set by equation $\tau=\tau_{c}>0$.

\section{Conclusion}
In conclusion the scope of problems associated with the
above-mentioned methods is briefly outlined.
\\
{\bf \\I.Involvement of Heisenberg's Algebra Deformation}
\\ One of the major problems associated with the proposed
approach to investigation of Quantum Mechanics of the Early
Universe  is an understanding of its relation to the Heisenberg' s
algebra deformation(e.g. see \cite{Magg}). It should be noted that
from the author's point of view the latter has two serious
disadvantages:
\\{\bf 1)} the deformation parameter is a dimensional
variable $\kappa$ with a dimension of mass;
\\{\bf 2)} in the limiting transition to QM this parameter goes
to infinity and fluctuations of other values are hardly sensitive
to it.
\\ At the same time, the merit of this approach is its ability
with particular assumptions to reproduce the Generalized
Uncertainty Relations.
\\ The proposed approach is free from such limitations as
{\bf 1)} and {\bf 2)}, since the deformation parameter is
represented by the dimensionless quantity $\alpha$ and the
variation interval $\alpha$ is finite $0<\alpha\leq1/4$. However,
it provides no direct reproduction of the Generalized Uncertainty
Relations. This approach is applicable in the general cases of
Quantum Mechanics with Fundamental Length irrespective of the fact
whether it is derived from the Generalized Uncertainty Relations
or in some other way.
\\ Because of this, involvement of the both approaches
in deformation of Quantum Mechanics is of particular importance.
\\{\bf \\II. The Approach as Applied to a Quantum Theory of Black Holes}
\\
{\bf 2.1 Bekenstein-Hawking formula strong derivation}
\\This paper presents certain results pertinent to the application of the
above methods in a Quantum Theory of Black Holes (subsections {\bf
4.2, 4.3}). Further investigations are still required in this
respect, specifically for the complete derivation of a
semiclassical Bekenstein-Hawking formula for the Black Hole
entropy, since in subsection 4.3 the treatment has been based on
the demonstrated result: a respective number of the degrees of
freedom is equal to $A$, where $A$ is the surface area of a black
hole measured in Planck's units of area $L_{p}^{2}$
(e.g.\cite{r14},\cite{r20}). Also it is essential to derive this
result from the basic principles given in this paper. Problems
{\bf 2.1} and {\bf 2.2} are related.
\\{\bf 2.2  Calculation of quantum corrections to the Bekenstein-Hawking
 formula}.
\\ In subsection 8.4 for the introduced logarithmic correction it
has been noted (see for example \cite{r22}) that it is coincident
with coefficient $a_{0}$  in formula (\ref{U28L}):
\\
$$Sp[\rho(\alpha)]-Sp^{2}[\rho(\alpha)]=\alpha+a_{0}\alpha^{2}
+a_{1}\alpha^{3}+... $$
\\
when using the exponential ansatz. It is clear that such a
coincidence is not accidental and further investigations are
required to elucidate this problem.
\\{\bf 2.3 Quantum mechanics and thermodynamics of black holes with
GUR}
\\Of interest is to consider the results of \cite{r17}, \cite{r31}
as related to the quantum-mechanical studies and thermodynamics of
black holes with GUR assumed valid rather than the Heisenberg
Uncertainty Relations. This is directly connected to the
above-mentioned problem of the associations between the density
matrix deformation considered in this work and Heisenberg's
algebra deformation.
\\
\\{\bf 2.4 Singularities and cosmic censorship hypothesis}
\\In subsection {\bf 4.2} a slight recourse has been made to the
case when Schwarzshild radius is $r=0$ that is associated with
going to value $\alpha=1$ and finally to a complex value of the
density pro-matrix trace $Sp[\rho (\alpha)]$. It should be noted
that the problem of singularities is much more complex \cite{Sing}
and is presently treated both physically and mathematically. It
seems interesting to establish the involvement of the results
obtained by the author in solving of this problem.
\\
\\{\bf III. Divergence in Quantum Field Theory}
\\It is obvious that once the fundamental length is included into a
Quantum Theory, ultra-violet (UV) divergences should be excluded
due to the presence of a maximum momentum determining the cut-off
\cite{r1}. In case under study this is indicated by the presence
of an exponential ansatz (subsection {\bf 3.2}). Note, however,
that for any particular theory it is essential to derive the
results from the basic principles with high accuracy and in good
agreement with the already available ones and with the
experimental data of QFT for the UV region without renormalization
\cite{Div1,Div2}.
\\{\bf IV. The Approach as Applied to Inflation Cosmology}
As we concern ourselves with the Early Universe (Planck's
energies), the proposed methods may be applied in studies of
inflation cosmology \cite{r10,r27,Inf}, especially as
Wheeler-DeWitt Wave Function of the Universe $\Psi$ \cite{r11} is
reliably applicable in a semiclassical approximation only
\cite{Vil}. The problem is formulated as follows: on what
conditions and in what way the density pro-matrix $\rho (\alpha)$
or its respective modification may be a substitute for $\Psi$ in
inflation models?
\\
\\{\bf V. High-Energy Deformation of Gravitation}
\\Since this work actually presents a study of physics at Planck's
scales, it is expedient to consider quantum-gravitational effects
which should be incorporated for specific energies. As a
development of the proposed approach this means the construction
of an adequate deformation of the General Relativity including
parameter $\alpha$, i.e. deformation of Einstein's Equations and
the associated Lagrangian involving parameter $\alpha$. Then the
question arises: and what about the space-time quantization? The
author holds the viewpoint that as the first approximation of a
quantized space-time one can use a portion of the Nonuniform
Lattice $Lat^{\tau}_{\widetilde{\alpha}}$ described in section 8
that is associated with small-number nodes or with high-valued
parameters $\widetilde{\alpha}$ and $\tau$ ,just which are used to
define the physics at Planck scale where the quantum-gravitational
effects are considerable. In this approximation for the prototype
of a point in the General Relativity may be taken an elementary
cell, i.e. as an element of the above-mentioned lattice with
small-number neighboring nodes. Then the associated deformation of
Einstein's should be considered exactly in this cell.
\\ Note that this section involves all the problems considered in I-IV.
\\ In summary it might be well to make three general remarks.
\\
\\{\bf 1)} It should be noted that in some well-known papers on GUR and
Quantum Gravity (e.g. see \cite{r1,r3,r5,Magg}) there is no
mention of any measuring procedure. However, it is clear that this
question is crucial and it cannot be ignored or passed over in
silence. We would like to remark that the measuring rule used in
\cite{r29}, (formula (5)) is identical to the ours.  In this paper
the proposed measuring rule (\ref{U6}) is a good initial
approximation to the exact measuring procedure of QMFL.
Corrections to this procedure could be defined by an adequate and
fully established description of the space-time foam (see
\cite{r25},\cite{Foam}) at Planck's scale.
\\
\\{\bf 2)} One of the principal issues of the present work is the
development of a unified approach to study all the available
quantum theories without exception owing to the proposed small
dimensionless deformation parameter $\widetilde{\alpha}_{\tau}\in
Lat^{\tau}_{\widetilde{\alpha}}$ that is in turn dependent on all
the fundamental constants $G,c,\hbar$ and  $k$.
\\ Thus, there is  reason to believe that lattices
$Lat_{\widetilde{\alpha}}$ and $Lat^{\tau}_{\widetilde{\alpha}}$
may be a universal means to study different quantum theories. This
poses a number of intriguing problems:
\\{\bf (1)} description of a set of lattice symmetries
$Lat_{\widetilde{\alpha}}$ and $Lat^{\tau}_{\widetilde{\alpha}}$;
\\ {\bf (2)} for each of the well-known physical theories
($\varphi^{4}$,QED,QCD and so on) definition of the selected
(special) points (phase transitions, different symmetry violations
and so on) associated with the above-mentioned lattices.
\\{\bf 3)} As it was noted in \cite{Fadd},
advancement of a new physical theory implies the introduction of a
new parameter and deformation of the precedent theory by this
parameter. In essence, all these deformation parameters are
fundamental constants: $G$, $c$ and $\hbar$ (more exactly in
\cite{Fadd} $1/c$ is used instead of $c$). As follows from the
above results, in the problem from \cite{Fadd} one may
redetermine, whether a theory we are seeking is the theory with
the fundamental length involving these three parameters by
definition: $L_{p}=\sqrt\frac{G\hbar}{c^3}$. Notice also that the
deformation introduced in this paper is stable in the sense
indicated in \cite{Fadd}.
\\
\\{\bf Acknowledgements}
\\The author would like to acknowledge Prof. Nikolai Shumeiko,
Director of the Belarusian National Center of Particles and High-
Energy Physics, and Dr. Julia  Fedotova,Scientific Secretary of
the Center, for their assistance contributing to realization of my
research plans and activities; Ludmila Kovalenko for her
assistance in editing and Sofia Titovich for her help in
preparation of this manuscript; Prof.D.V.Ahluwalia-Khalilova,
Center for Studies of Physical, Mathematical and Biological
Structure of Universe, Department of Mathematics, University of
Zacatecas, Mexico for her interest in the subject matter; Profs.
Sergei Kilin, Vassilii Strazhev and also Dr. Arthur Tregubovich
for valuable discussions and remarks; and last but not the least
my wife Nadya Anosova for the support and encouragement when
working on this chapter.


\end{document}